\documentclass[aps, twocolumn, showpacs, superscriptaddress, floatfix]{revtex4-2}

\usepackage[utf8]{inputenc}
\usepackage{graphicx}
\usepackage{dcolumn}
\usepackage{bm}
\usepackage[utf8]{inputenc}
\usepackage[T1]{fontenc}
\usepackage{mathptmx}
\usepackage{gensymb}
\usepackage{upgreek}
\usepackage{xcolor}
\usepackage{textcomp}

\usepackage{hyperref}
\hypersetup{colorlinks=true,linkcolor=blue,citecolor=blue,filecolor=magenta,urlcolor=magenta}
\usepackage{url}
\usepackage{float}

\newcommand{\reffig}[2]{Fig.{\color{blue} \,\ref{#1}#2}}
\newcommand{\cmt}[1]{{}}
\usepackage{titlesec}
\titleformat{\section}[hang]{\normalfont\bfseries\sffamily}{\thesection.}{0.5em}{}
\titlespacing{\section}{0pc}{1.2pc}{0.3pc}

\newcommand{\rom}[1]
    {\MakeUppercase{\romannumeral #1}}

\begin{document}

\title{Photonic link from single flux quantum circuits to room temperature}

\author{Mohan Shen}
\affiliation{Department of Electrical Engineering, Yale University, New Haven, CT 06511, USA}
\author{Jiacheng Xie}
\affiliation{Department of Electrical Engineering, Yale University, New Haven, CT 06511, USA}
\author{Yuntao Xu}
\affiliation{Department of Electrical Engineering, Yale University, New Haven, CT 06511, USA}
\author{Sihao Wang}
\affiliation{Department of Electrical Engineering, Yale University, New Haven, CT 06511, USA}
\author{Risheng Cheng}
\affiliation{Department of Electrical Engineering, Yale University, New Haven, CT 06511, USA}
\author{Wei Fu}
\affiliation{Department of Electrical Engineering, Yale University, New Haven, CT 06511, USA}
\author{Yiyu Zhou}
\affiliation{Department of Electrical Engineering, Yale University, New Haven, CT 06511, USA}
\author{Hong X. Tang}
\email{hong.tang@yale.edu}
\affiliation{Department of Electrical Engineering, Yale University, New Haven, CT 06511, USA}

\date{\today}

\begin{abstract}

Broadband, energy-efficient signal transfer between cryogenic and room-temperature environment has been a major bottleneck for superconducting quantum and classical logic circuits. Photonic links promise to overcome this challenge by offering simultaneous high bandwidth and low thermal load. However, the development of cryogenic electro-optic modulators --- a key component for photonic readout of electrical signals --- has been stifled by the stringent requirements of superconducting circuits. Rapid single flux quantum circuits (RSFQ), for example, operate with a tiny signal amplitude of only a few millivolts (mV), far below the volt-level signal used in conventional circuits. Here, we demonstrate the first direct optical readout of an RSFQ circuit without additional electrical amplification enabled by a novel superconducting electro-optic modulator (SEOM) featuring a record-low half-wave voltage $V_{\rm{\pi}}$ of 42\,mV on a 1\,m-long SEOM. Leveraging the low ohmic loss of superconductors, we break the fundamental $V_{\rm{\pi}}$-bandwidth trade-off and demonstrate electro-optic bandwidth up to 17\,GHz on a 0.2\,m-long SEOM at cryogenic temperatures. Our work presents a viable solution toward high-bandwidth signal transfer between future large-scale superconducting circuits and room-temperature electronics.

\end{abstract}
\maketitle

Superconducting circuits are among the most promising technologies for quantum information processing\cite{Devoret2013, Arute2019} and ultra-fast logic circuits \cite{Holmes2013, Braginski2019}. Fulfilling the superiority these cryogenic computation schemes promise, whether classical or quantum, relies on the development toward large-scale superconducting integrated circuits (ICs) \cite{Das2017, Fowler2012}. A fundamental roadblock on this scaling roadmap is their connectivity to room-temperature electronics, which has so far relied on coaxial cables that have limited bandwidth and finite thermal conductivity \cite{Krinner2019}. Also, to sustain the ultra-low-level signal emerging from the superconducting circuits within a coaxial cable, multi-stage amplifications are needed at cryogenic temperatures. These amplifiers add significant thermal load to the superconducting ICs and overall cryogenic cooling budget \cite{Krinner2019}. To address these challenges, photonic links using optical fibers have been identified as a promising solution \cite{Lecocq2021a, Youssefi2020, DeCea2020b, Han2021}. Compared to electrical cables, optical fibers offer two orders of magnitude lower heat load and three orders of magnitude higher bandwidth \cite{DeCea2020b}. Additionally, they are less susceptible to thermal noise and crosstalk. Optical data links at room temperature also enable broadband and low-loss data transportation over local data centers \cite{Miller2017, Kachris2012} and across remote networks \cite{Winzer2018}.

Similar to the room-temperature fiber links, the implementation of cryogenic-to-room-temperature photonic links critically relies on electro-optic (EO) modulators to transduce microwave signals into the optical domain. Superconducting circuits, however, impose significantly more demanding requirements on the modulators than their room-temperature counterparts, not only in terms of stringent cryogenic compatibility but also in terms of infinitesimal signals to be uplifted. One important class of superconducting logic ICs is the single flux quantum (SFQ) logic family for digital signal processing, which has been utilized in radiofrequency (RF)-digital receivers, development of next-generation energy-efficient computers and proposed for large-scale control and readout of superconducting qubits \cite{MUKHANOV2008,Mukhanov2011,McDermott2018}. These Josephson-junction-based circuits encode the digital information in quantized magnetic flux, which allows operations with atto-Joules of energy and ultra-fast switching above tens of gigahertz, such as rapid single flux quantum (RSFQ) devices \cite{Likharev1991a}. Since the signal generated by SFQ circuits is only a few millivolts (mV) in amplitude, readout of this signal has so far relied on additional electrical amplification to hundreds of mV \cite{Gupta2019a, Fu2022}. Direct photonic links to SFQ circuits have not been realized so far as it requires a broadband EO modulator capable of mV-scale operations.

Another important class of superconducting circuits is that for quantum information processing, where the uplift of quantum states to room temperature calls for efficient microwave-to-optics quantum transduction \cite{Lambert2020a, Han2021}. To achieve enhanced conversion efficiency, cavity-based structures are universally utilized, but at the sacrifice of conversion bandwidth. A recent work \cite{Delaney2022} demonstrated optical readout of a superconducting qubit with low back-action using an EO transducer with around $10^{-3}$ transduction efficiency, nonetheless with a transduction bandwidth of only a few kilohertz. With a high-bandwidth EO modulator of similar or moderately improved transduction efficiency, a fully photonic interface for multiplexed readout (egress) of large-scale superconducting qubits can be envisioned \cite{Usami2021}, which complements recent photonic (ingress) links demonstrated for transmitting qubit control signals from room temperature to cryogenic environment \cite{Lecocq2021a}.

Cryogenic modulators have recently been demonstrated on different platforms with volt-scale modulation amplitudes \cite{Gehl2017, Eltes2020, Chakraborty2020, Lee2020, Yin2021, Pintus2022a}. Leveraging cavity enhancement, a recent semiconductor-based microring modulator achieved a 10\,mV drive voltage with a 3\,dB bandwidth close to 1\,GHz \cite{Pintus2022}. Similar to resonant microwave-to-optics converters, the bandwidth of microring-based modulators is ultimately limited by the cavity linewidth \cite{Xiong2012a}. On the other hand, traveling-wave EO modulators have been widely used in telecommunication networks because of their high bandwidth \cite{Wooten2000}, such as lithium niobate (LN)-based traveling-wave modulators. Studies showed the compatibility of LN modulators with cryogenic operating environment \cite{Herzog2008a, Lomonte2021b} and demonstrated proof-of-principle cryogenic optical interconnect using a commercial LN EO modulator \cite{Youssefi2020}. However, the large $V_{\rm{\pi}}$ ($\sim 5\,\mathrm{V}$) of these commercial EO modulators leads to a low transduction efficiency ($3.5\times 10^{-7}$), thus making them less competitive with HEMT-amplified electrical links. Here, $V_{\rm{\pi}}$ is the voltage needed to introduce a $\pi$ phase shift on the modulator's optical output and can be translated to the modulator's transduction efficiency (see Supplementary Information Sec.\,\rom{3}). Recent advances in integrated modulators based on lithium niobate on insulator (LNOI) platform \cite{Wang2018} have reduced $V_{\rm{\pi}}$ to 1-2\,V, which however still falls short of the demanding requirements of cryogenic applications. For example, reaching a $10^{-3}$ transduction efficiency for superconducting qubit readout necessitates a modulator with $V_{\rm{\pi}}$ in the range of 100-200\,mV (Supplementary Information Sec.\,\rom{3}).

\begin{figure}
    \centering
    \includegraphics{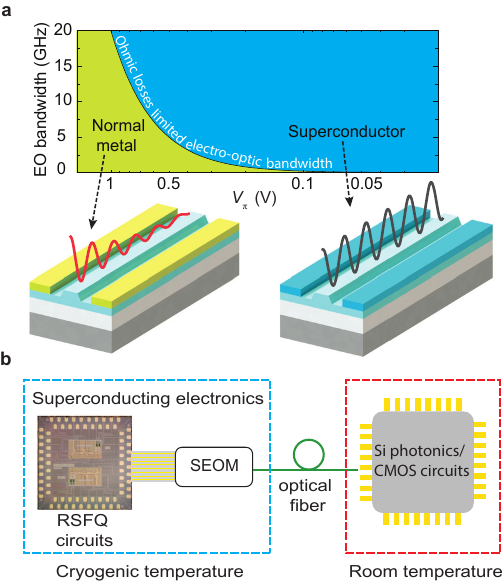}
    \caption{Superconducting electro-optic modulator (SEOM) and its application in cryogenic-to-room-temperature link. \textbf{a}, The microwave loss in normal metal leads to a fundamental trade-off between modulator's bandwidth and its V$_\pi$. SEOM breaks this limit and promises a vastly expanded parameter space. For the normal metal microwave propagation loss we use a typical value of $\alpha= 0.7\,\mathrm{dB\,cm^{-1}}{(f/\mathrm{Hz})}^{1/2}$
    \cite{Zhang2021} and thus $f_{\rm{3dB}}=20\,\mathrm{GHz}(V_\pi/\mathrm{V})^{2}$.
    \textbf{b}, Illustration of an SEOM enabled photonic link between superconducting and room-temperature electronics.}
    \label{fig:fig1}
\end{figure}

A straightforward approach to reduce the $V_{\rm{\pi}}$ of traveling-wave modulators and improve their transduction efficiency is to increase the modulation length from the current centimeter range to the decimeter or even meter range. On top of the technical challenges of fabricating extremely long modulators, there exists a fundamental limit between bandwidth and efficiency on traveling-wave modulator architecture, imposed by the RF attenuation from ohmic losses \cite{Zhang2021}. Taking the typical electrode design of an integrated LN EO modulator, the ohmic-loss-limited bandwidth can be estimated as $f_{\rm{3dB}}=20\,\mathrm{GHz}\,{(V_\mathrm{\pi}/\mathrm{V})^{2}}$ \cite{Zhang2021}. As illustrated in \reffig{fig:fig1}{a}, this limitation severely restricts the performance parameter space of the modulator, particularly with low-$V_{\rm{\pi}}$ devices. To overcome this limit, researchers postulated modulators with superconductor electrodes and have demonstrated that low-loss superconducting microwave transmission line could be employed to increase the effective modulation length \cite{Yoshida1999}. This idea has recently resurfaced due to the recently heightened interests in ultra-low $V_{\rm{\pi}}$ and high-bandwidth cryogenic modulators \cite{Youssefi2020}. In the previous work \cite{Yoshida1999}, although superconductor electrodes were employed in the modulator, the modulation length is only 2\,cm ($V_\pi\sim4\,\mathrm{V}$) and the device performance was not extended to the regime where the superconductor is more competitive than normal metals.

In this Article, we demonstrate the concept of superconducting traveling-wave modulators based on thin-film LN for interfacing with superconducting logic circuits. This superconducting electro-optic modulator (SEOM) design combines the best of two worlds in terms of material performance --- the low microwave loss of superconductors and the low optical loss / high EO coefficient of LN --- thus drastically increasing the EO modulation efficiency. Through an electrode jump-over design, the modulation length can be extended to one meter long (0.5\,m in each Mach-Zehnder arm) while still maintaining a compact footprint, thus reducing the $V_{\rm{\pi}}$ to as low as 42\,mV. We also show that at cryogenic temperatures, the superconducting modulator possesses a 3\,dB bandwidth over 17\,GHz (20\,cm total modulation length) by matching the velocity of microwave and optical signals, in sharp contrast to the modulation bandwidth of a few megahertz when the electrodes are normal. Our results suggest that SEOM can break the fundamental trade-off between $V_{\rm{\pi}}$ and modulation bandwidth in EO modulator designs. We further demonstrate cryogenic-to-room-temperature data link with low peak-to-peak voltage ($V_{\rm{pp}}$) and achieve direct data lifting from an RSFQ circuit (5\,mV $V_{\rm{pp}}$)\cmt{at 1\,Gbps}. We believe our superconducting modulator design provides a pathway toward future high-bandwidth optical link for cryogenic integrated circuits.

\begin{figure*}
    \centering
    \includegraphics{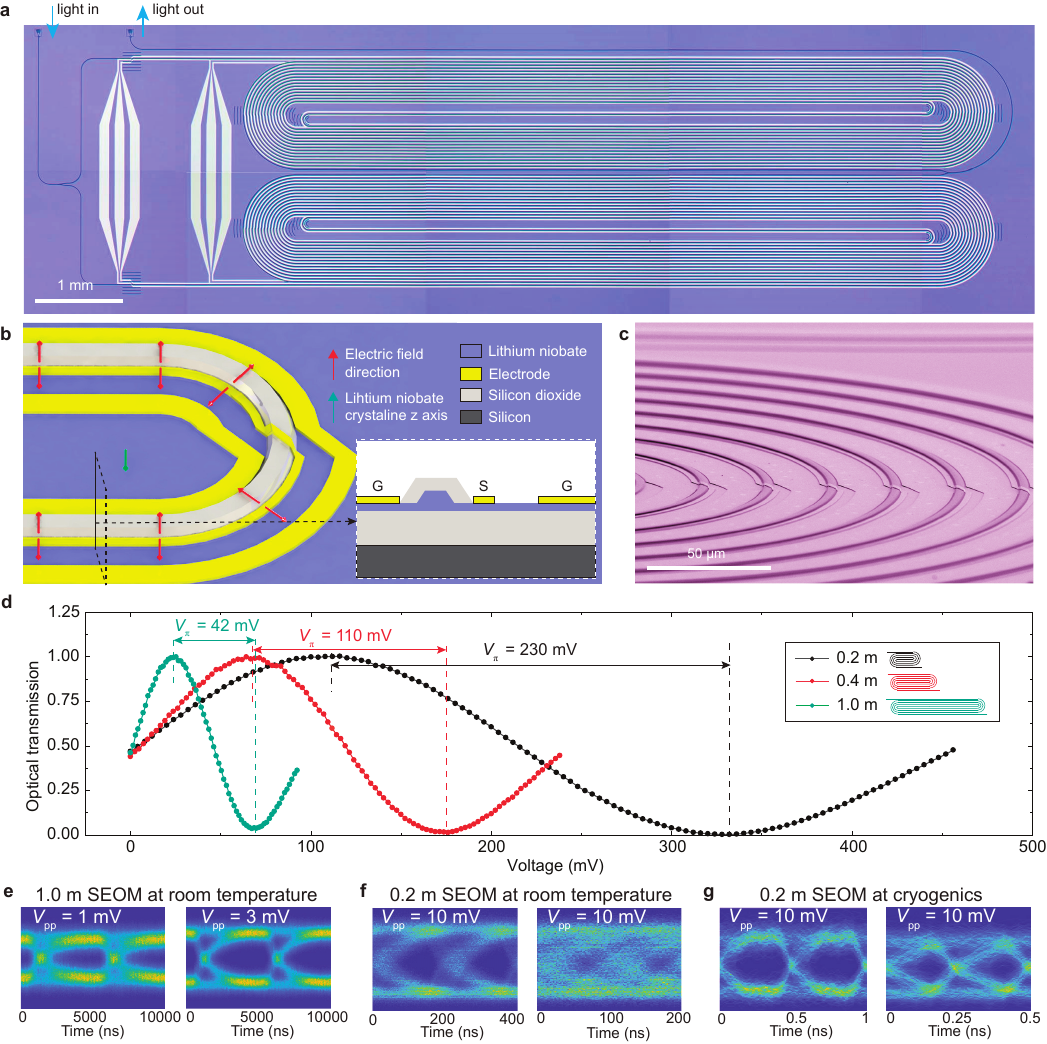}
    \caption{
    Low-drive-voltage operation of SEOM.
    \textbf{a}, Optical micrograph of a 0.4-meter-long (0.2\,m each arm) SEOM device. 
    \textbf{b}, Schematic illustration of the jump-over superconductor electrodes. By allowing the electrodes to cross over the waveguide, this design re-orients the modulation electric field as the optical mode is looped back to prevent cancellation of the modulation effect. Inset shows the modulator cross-section.
    \textbf{c}, SEM image of the jump-over structures.
    \textbf{d}, Room temperature $V_{\pi}$ measurement. The measured $V_{\rm{\pi}}$ of three SEOM with total modulation length of 0.2\,m, 0.4\,m and 1.0\,m is 230\,mV, 110\,mV and 42\,mV respectively. 
    \textbf{e-g}, Eye diagrams. In \textbf{e}, eye diagrams for 1\,mV and 3\,mV drive voltage are demonstrated with the 1\,m SEOM at room temperature. In \textbf{f}, the limited bandwidth due to normal metal resistance is manifested by the diminishing eye opening as the bit rates are increased. While for SEOM, this bandwidth limit is lifted and eye diagrams with 10\,mV drive voltage at 2\,Gbps and 4\,Gbps are demonstrated in \textbf{g}. The eye diagrams are taken by a 6\,GHz oscilloscope.
    }
    \label{fig:fig2}
\end{figure*}
\vskip6pt

\section*{Results}

\noindent\textbf{Ultra-low $V_{\rm{\pi}}$.} For a specific EO material, the strategy to reduce $V_{\rm{\pi}}$ is to fabricate modulators with extended modulation length. Here, we present our design to create the longest LN modulator to date, thereby achieving the lowest $V_{\rm{\pi}}$ ever reported. The devices are fabricated on an $x$-cut LNOI wafer with a 600-nm-thick LN film. In the two arms of the Mach–Zehnder interferometer (MZI) modulator (\reffig{fig:fig2}{a}), the long waveguide is laid out spirally with the extended straight sections aligned along the crystalline $y$-axis to harness $r_{\rm{33}}$, the largest EO coefficient of LN. A 25\,$\mathrm{\micro m}$ waveguide spacing is designed to accommodate a ground-signal-ground (GSG) microwave transmission line in between. The input microwave signal splits at the input port and propagates along each spiral arm of the MZI, and is terminated at the output port. With this design, a total 0.4\,m-long optical waveguides can be fit into a 10\,mm by 2.5\,mm area (14\,mm by 4\,mm for a 1\,m-long modulator). Due to the large dielectric constant of LN, the optical waveguide is partially etched so that the electrodes are deposited directly on the slab to enhance the electro-optic mode overlap, as shown in \reffig{fig:fig2}{b} inset. Niobium (Nb) is chosen as the electrode material for its high superconducting transition temperature. Although there are other superconductor materials with higher transition temperatures, these materials are typically compounds and their high kinetic inductance makes the velocity matching between optical and microwave signals challenging. A more quantitatively analysis on microwave transmission line design is provided in the next section.

Note that a critical enabling design for the microwave transmission line is the electrode jump-over structures where the signal electrode climbs over the optical waveguide from one side to the other at each waveguide bending. In this way, the modulation electric field always orients in the same direction as the optical mode propagates along the meander, as illustrated by the red arrows in \reffig{fig:fig2}{b}. Without the jump-over structures, the modulation effect cancels out. To minimize optical loss induced by metal absorption at the jump-over structure, we clad the waveguide with dielectric that retains a slant sidewall to ensure  electrical continuity of the electrodes. \reffig{fig:fig2}{c} shows the scanning-electron microscopy (SEM) picture of the electrode jump-over structures.

With the design presented above, we are able to fabricate modulators with up to 1\,m-long modulation length. The $V_{\rm{\pi}}$ of the fabricated MZI modulators with total modulation length of 0.2\,m, 0.4\,m and 1\,m (0.1\,m, 0.2\,m and 0.5\,m in each arm) are measured to be 230\,mV, 110\,mV and 42\,mV respectively at room temperature, as shown in \reffig{fig:fig2}{d}. Following the convention to determine the voltage-length product ($V_\pi L$) using the modulation length of one arm, we measure the $V_\pi L$ value to be around $2.2\,\mathrm{V \cdot cm}$. At 4\,K, we observe a 70\% increase on $V_{\rm{\pi}}$ (data included in Supplementary Information Sec.\,\rom{5} ). The $V_{\rm{\pi}}$ change of LN based modulators at cryogenic temperatures has been reported in different studies, and the reported value varies from a decrease of 10\% \cite{Youssefi2020} to an increase of 20\% \cite{Herzog2008a} and 74\% \cite{Thiele2022}. We think the $V_{\rm{\pi}}$ change is resulted from the temperature dependence of the electro-optic coefficient. The discrepancy on the reported results might be due to different material growth methods and fabrication conditions. With some of the reported results showing that the $V_{\rm{\pi}}$ decreases or does not substantially change, it is possible that the $V_{\rm{\pi}}$ increase could be avoided, but this is subject to future studies.

The ultra-low $V_{\rm{\pi}}$ allows for mV-drive-voltage operation. At room temperature, we demonstrate eye diagrams with $V_{\rm{pp}}$ as low as 3\,mV and 1\,mV using a 1\,m-long modulator, as shown in \reffig{fig:fig2}{d}. However, the modulation speed is only 200\,kbps, limited by the large resistance of the long normal metal electrodes. Here the dimension of the electrode is intentionally kept narrow (1-2\,$\mathrm{\micro m}$ in width and 200-300\,nm in thickness for the signal electrode) compared to conventional modulator design in order to reduce the device overall footprint. This results in a room-temperature electrode resistance on the order of 1\,k$\Omega$/cm, which sets the modulator bandwidth limited by the resistance-capacitance (RC) time constant\cmt{ in the lumped element regime }. The simulated specific capacitance of the microwave transmission line is around 1\,pC/cm, corresponding to a bandwidth of a few megahertz for the 0.2\,m-long (0.1\,m each arm) modulator. This bandwidth limit can be seen from the two eye diagram measurements with 10\,mV $V_{\rm{pp}}$ in \reffig{fig:fig2}{f}, where the eye closes when the modulation speed increases from 5\,Mbps to 10\,Mbps. This bandwidth limit is lifted after the electrodes turn superconducting. The last two eye diagrams in \reffig{fig:fig2}{g} shows 3 orders of magnitude larger bandwidth after the superconducting transition with 10\,mV $V_{\rm{pp}}$ at 2\,Gbps and 4\,Gbps. The eye diagram data is taken by an oscilloscope with 6\,GHz bandwidth. Although the EO analog bandwidth of our SEOM is above 17\,GHz (as shown in the next section), eye diagram operation at higher rate requires higher optical power to maintain the same signal-to-noise ratio (Supplementary Information Sec.\,\rom{7}). The operation rate is primarily constrained  by our current high optical insertion loss (20\,dB) and the associated optical heating effect. Further details regarding these limitations are discussed in the subsequent sections.

\begin{figure}
    \centering
    \includegraphics{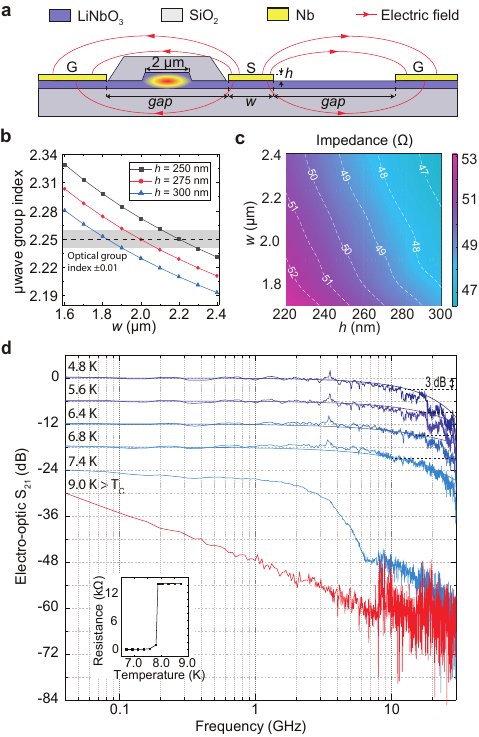}
    \caption{SEOM bandwidth: modeling and experimental characterization. 
    \textbf{a}, Cross-sectional model of the LN optical waveguide and superconductor microwave transmission line. The microwave transmission line has a GSG configuration. The LN ridge waveguide is cladded by $\mathrm{SiO_2}$ and placed between the signal and ground electrodes. $h$, $w$ and $gap$ denote the electrode thickness, signal electrode width and the gap between signal and ground electrodes.
    \textbf{b-c}, Simulated microwave group index and characteristic impedance. Fixing $gap = 5.2\,\mathrm{\micro m}$, the microwave group index can be matched to that of optics when the Nb film thickness is around 275\,nm and the signal electrode width is around 2\,$\mathrm{\micro m}$. Simultaneously, the transmission line impedance can be matched to 50\,$\Omega$ on this LNOI platform. 
    \textbf{d}, EO bandwidth measurement of a 0.2\,m SEOM at different temperatures. The measured EO responses at 4.8\,K, 5.6\,K, 6.4\,K and 6.8\,K are fitted (line), from which we derive a 3\,dB bandwidth of 13.5\,GHz, 17.5\,GHz, 16.8\,GHz and 11.0\,GHz respectively. When the temperature is above $T_c$, the bandwidth is dropped to tens of megahertz. The inset shows the DC resistance of the signal electrode of the transmission line drops from 13.8\,$\mathrm{k\Omega}$ to zero as the temperature decreases below $T_c$ of 8\,K.
    }
    \label{fig:fig3}
\end{figure}
\vskip6pt

\noindent \textbf{Electro-optic bandwidth of SEOM.} The modulation bandwidth of a traveling-wave EO modulator is determined by several factors: 1) the group velocity mismatch between the optical and microwave modes; 2) the propagation loss of each mode; 3) microwave dispersion control and impedance matching. A theoretical derivation of the bandwidth dependence on these factors is provided in the Supplementary Information Sec.\,\rom{6}. With negligible microwave loss and assuming low microwave dispersion (as suggested by simulation), the key to high EO bandwidth is the group velocity matching between the optics and microwave. A careful impedance engineering to match the on-chip transmission line impedance with the $50\,\Omega$ coaxial cable is also necessary for high-bandwidth operation.

With these considerations in mind, we model the microwave transmission line electro-optically coupled with the optical waveguide (\reffig{fig:fig3}{a}). The silicon dioxide layer between LN and silicon substrate is $4.7\,\mathrm{\micro m}$ thick. The optical waveguide geometry is $2\,\mathrm{\micro m}$ wide and 600\,nm thick with 250\,nm slab to maintain low optical propagation loss. The width of the signal electrode of the transmission line $w$, electrode thickness $h$ and the gap between electrodes $gap$ are swept to adjust the microwave transmission line speed and impedance. The gap between the electrodes is chosen to be $5.2\,\mathrm{\micro m}$ to minimize the metal absorption loss while maintain a relatively low $V_\pi L$ (Supplementary Information Sec.\,\rom{3}). The optical group index is simulated to be around 2.25 and the microwave group index can be effectively adjusted by the signal electrode width and thickness as shown in \reffig{fig:fig3}{b}. In the simulation, the kinetic inductance of the superconductor is taken into account through the equations in \cite{Baker1974}, where we assume a uniform current distribution in the signal electrode. With the Nb signal electrode of 275\,nm thickness and $2\,\mathrm{\micro m}$ width, the simulated specific capacitance, geometric and kinetic inductance are 0.74\,pC/cm, 6.2\,nH/cm and 1.2\,nH/cm, and the microwave index can be matched to that of optics and its impedance can be matched to 50\,$\Omega$ simultaneously (\reffig{fig:fig3}{c}). Note that low kinetic inductance enables the velocity and impedance match. This could also be achieved by using other elementary superconductors, like aluminum or indium, but it would be challenging for compounds superconductor like NbN, TiN or NbTiN, whose kinetic inductance can be one order of magnitude higher.

In \reffig{fig:fig3}{d}, we assess the EO bandwidth of a 0.2\,m SEOM at cryogenic temperatures as the device is cooled below the superconducting transition temperature $T_c$ of Nb electrodes ($\sim 8$\,K, \reffig{fig:fig3}{d} inset). Above $T_c$, the signal electrode has a resistance of 13.8\,k$\Omega$, which expectedly limits the EO bandwidth to tens of megahertz, as shown in the EO response measured at 9\,K. After the electrodes turns superconducting, the EO bandwidth immediately expands by more than three orders of magnitude. Although our simulations predict that the velocity and impedance matching can be simultaneously fulfilled on the LNOI platform, the actual device performance might differ from the simulations. This potential deviation could be due to the fabrication imperfections and the uncertainties of materials' coefficients that we incorporated in our simulations. In experiments, this deviation can be compensated by fine tuning of kinetic inductance of the superconductor electrodes \cite{tinkham2004introduction}. The temperature dependence of the microwave index and propagation loss are measured and included in Supplementary Information Sec.\,\rom{6}. With these measured results, we fit the response curves using the theoretical model depicted in Supplementary Information Sec.\,\rom{3}. As shown in \reffig{fig:fig3}{d}, the highest EO bandwidth of 17.5\,GHz is achieved at 5.6\,K. According to our model and fitting, the index is best matched at around 6.4\,K, but as this temperature is approaching $T_c$, the higher microwave loss limits the EO bandwidth (details in Supplementary Information Sec.\,\rom{6}). For future devices, achieving index matching at a lower temperature with lower microwave loss will further improve the EO bandwidth.

\begin{figure*}
    \centering
    \includegraphics{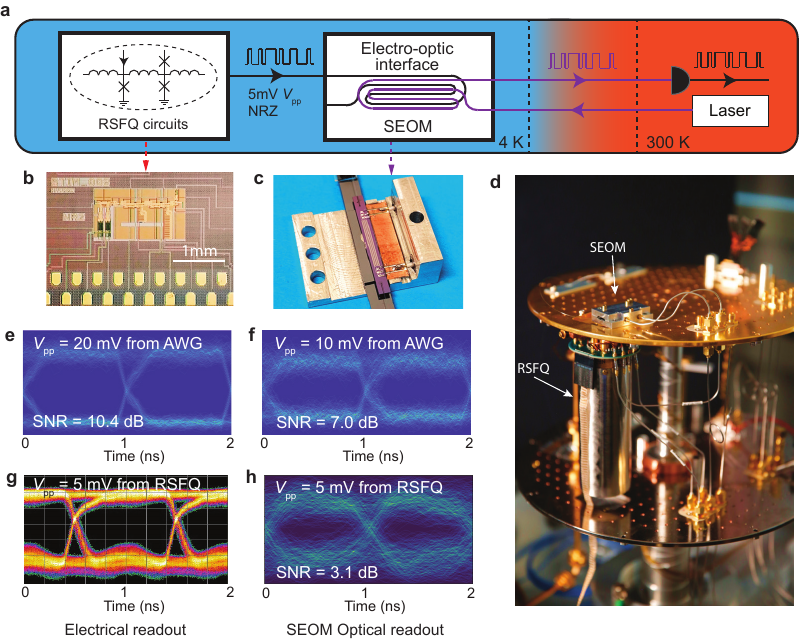}
    \caption{Photonic link from an RSFQ circuit to room temperature. 
    \textbf{a}, Schematic illustration of the photonic link interfacing RSFQ circuits to room temperature through SEOM. 
    \textbf{b}, Micrograph of the RSFQ IC. 
    \textbf{c}, Packaged SEOM device. 
    \textbf{d}, Picture showing the physical interface between RSFQ and SEOM. The RSFQ module is packaged in a mu-metal shield and mounted to the back side of the sample plate of a 4\,K cryostat whereas the SEOM module is mounted on the top side. The output signal from the RSFQ is directly routed to the SEOM through coaxial cables without intermediate amplification. 
    \textbf{e-f}, Eye diagrams generated by the packaged SEOM when driven by an arbitrary waveform generator (AWG) at room temperature. The PRBS7 data stream in NRZ format is sent to the SEOM with peak-to-peak amplitude of 20\,mV and 10\,mV at 1\,Gbps. The SNR is 10.4\,dB and 7.0\,dB for 20\,mV and 10\,mV drive voltage. 
    \textbf{g}, Output signal of the RSFQ circuit characterization. The PRBS7 NRZ signal generated by the RSFQ circuit is amplified by 25\,dB at 40\,K and subsequently routed to room temperature. 
    \textbf{h}, Direct optical readout of the 5\,mV $V_{\rm{pp}}$ RSFQ signal. The SNR of the photodetected signal is 3.1\,dB.
    }
    \label{fig:fig4}
\end{figure*}

 \noindent \textbf{Optical readout of an RSFQ circuit.} Enabled by the ultra-low $V_{\rm{\pi}}$ and high bandwidth of our SEOM, we demonstrate the first direct optical readout of an RSFQ circuit. The RSFQ chip is designed by HYPRES, Inc and fabricated at MIT Lincoln Laboratory using SFQ5EE process \cite{Tolpygo2016}. It employs an on-chip SQUID-stack driver to output a 5\,mV $V_{\rm{pp}}$ signal into 50\,$\Omega$ impedance up to 10\,Gbps. The RSFQ chip is packaged in a mu-metal housing mounted on the back side of the sample plate of a 4\,K cryostat. To interface with the RSFQ circuit, our SEOM is fully packaged to provide  modularized EO interface. As shown in \reffig{fig:fig4}{c}, the on-chip electrode pads are wire-bonded to a printed circuit board (PCB) with very short wires (1-2\,mm) to ensure a high bandwidth connection. The PCB is then connectorized on the aluminum housing. Photonic packaging is through on-chip optical grating coupler coupled to single-mode optical fibers (Supplementary Information Sec.\,\rom{8}). The SEOM module (a 0.2\,m SEOM is used in this experiment, $V_\pi$=220\,mV @ 300\,K and 380\,mV @ 4\,K) is mounted on the top side of the sample plate and directly connected to the RSFQ module through coaxial cables, as shown in \reffig{fig:fig4}{d}. Therefore, the output data stream from the RSFQ circuit is first routed to the SEOM device, where it is translated into optical domain and subsequently lifted to room temperature through optical fibers.

To validate the SEOM's EO performance at low excitation signal levels, we first send signals from an arbitrary waveform generator (AWG) at room temperature through different attenuations to drive the modulator. With 20\,mV and 10\,mV peak-to-peak pseudo-random bit sequences (PRBS7) in non-return-to-zero (NRZ) format at 1\,Gbps, the packaged device generates clear eye diagrams as shown in \reffig{fig:fig4}{e-f}, suggesting that this SEOM device can directly handle the data stream emerging from the RSFQ circuit, which outputs signals with a typical $V_{\rm{pp}}$ of about 5\,mV. 

Next, we switch to the RSFQ circuit as signal source. To electrically visualize the weak signal generated by RFSQ, we have to amplify the signal at cryogenic temperatures, which would otherwise be submerged by room-temperature noise. The electrical readout of the PRBS7 NRZ signal uses an amplifier with 35\,dB gain at 4\,K and the signal is shown in \reffig{fig:fig4}{g}. Photodetected signal after SEOM without any electrical amplification is displayed in \reffig{fig:fig4}{h}. With the decreasing drive voltage from 20\,mV, 10\,mV to 5\,mV, the signal-to-noise ratio (SNR) degrades from 10.4\,dB, 7.0\,dB to 3.1\,dB. The corresponding bit error rates (BER) are calculated to be $1.4\times10^{-6}$, $7.7\times10^{-4}$ and $2.2\times10^{-2}$ respectively \cite{Freude2012}. The BER of $2.2\times10^{-2}$ is within the tolerance of forward error correction (20\% overhead with tolerable BER of $2.7\times10^{-2}$ \cite{Agrell2018, chang2012ldpc}). Here, a significant limitation on the SNR arises from the high optical insertion loss at cryogenic temperatures. The total optical loss is 20\,dB, which mostly comprises a 12\,dB coupling loss and an additional 8\,dB on-chip insertion loss. To compensate for the optical loss, we use a relatively high laser power of 17\,dBm for the SEOM device. At this input power level, the calculated transduction efficiency is $1.4\times 10^{-4}$ as shown in \reffig{fig:fig5}{}. The current SNR limitation hinders our ability to demonstrate higher-speed eye diagrams, despite the SEOM device's inherent high bandwidth (Supplementary Information Sec.\,\rom{7}). In the following section, we will explore into this SNR limitation and highlight the potential of enabling optical readout of the RSFQ circuit at 10\,Gbps with significantly lower optical power by further reducing optical losses.

\section*{Discussion}

\begin{figure}
    \centering
    \includegraphics{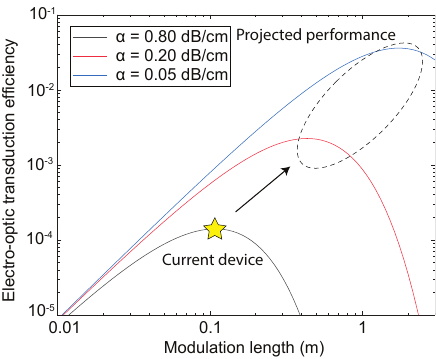}
    \caption{Projected SEOM transduction efficiency. The electro-optic transduction efficiency of SEOM is calculated as a function of modulation length under different optical propagation loss conditions. The calculation assumes a 10\,dBm optical input power and uses a cryogenic $V_\pi L$ of $3.8\,\mathrm{V \cdot cm}$ given by experimental results. The current SEOM device possesses a transduction efficiency of $1.4\times 10^{-4}$ under 0.8\,dB/cm propagation loss. With improved propagation loss to 0.2\,dB/cm and 0.05\,dB/cm, transduction efficiency can approach 1\% and 10\% with meter-long modulation length. }
    \label{fig:fig5}
\end{figure}

\noindent We have established that the use of superconductors breaks the trade-off between $V_{\rm{\pi}}$ and modulation bandwidth in EO modulator designs by suppressing the ohmic loss. Nevertheless, the optical loss comes into play as the limiting factor for meter-long modulators. As shown in Supplementary Information Sec.\,\rom{3}, the transduction efficiency increases quadratically with the length of modulation, while the light intensity decays exponentially as it propagates. With a given optical input power, the transduction efficiency $\eta$ (or equivalently the photon number per bit in eye diagram measurement) is given by:
\begin{equation}
   \eta = 
    P_{\mathrm{opt}}\left[ {\tiny \frac{\pi^2}{2} \frac{\Omega}{\omega} \frac{Z_0}{(V_\pi L)^2}} \right] e^{-\alpha_\mathrm{o} L}L^2,
     \label{eq:eq1}
\end{equation}
where $P_\mathrm{opt}$ is the optical input power, $\Omega$ and $\omega$ are the microwave and optical angular frequency, $\alpha_\mathrm{o}$ is the optical propagation loss, $Z_0$ is the microwave transmission line characteristic impedance, and $L$ is the modulation length. As the voltage-length product $V_\pi L$ is a constant, Eq.\,(\ref{eq:eq1}) has only one independent variable $L$ and it implies that there is an optimal modulation length for a given propagation loss $\alpha_o$ ($L=2/\alpha_o$). Our current device still experiences residual photorefractive loss which leads to a propagation loss of about 0.8\,dB/cm (Supplementary Information Sec.\,\rom{9}). This material degradation can be traced to unrepaired damage induced by electron-beam exposure during our lithography process, and we can not recover this damage through the conventional thermal annealing process because the superconductor cannot survive the high annealing temperature. This fabrication-induced damage could be avoided or recovered, for example, by changing our current e-beam lithography to photo-lithography or performing the annealing at an ultra-high vacuum chamber. Without the material damage, 0.027\,dB/cm has been reported on microring resonators \cite{Zhang2017b}. Given the current 0.8\,dB/cm propagation loss, the optimal modulation length is around 10\,cm in each MZI arm, as shown in \reffig{fig:fig5}{a}. With recovered material damage and thus a lower propagation loss, the transduction efficiency can be further improved at a longer modulation length. With 0.05\,dB/cm propagation loss, the transduction efficiency will be above 1\%.

Our current cryogenic power dissipation is also limited by the large optical losses. The power dissipation is composed of two parts: electrical and optical power dissipation. Electrically, power only dissipates at the end of the microwave transmission line when it is terminated by a 50\,$\Omega$ load. Although a DC voltage bias is applied between the signal and ground electrodes, this pure voltage bias on the dielectric material does not consume energy. The dissipated microwave power can be calculated as $V^2/Z_0$, where $V$ is the signal amplitude (one half of $V_{\rm{pp}}$) and $Z_0$ is the microwave impedance. In our RSFQ optical readout demonstration, the 5\,mV $V_{\rm{pp}}$ signal at 1\,Gbps corresponds to 125\,aJ/bit. If operated at a higher rate of 10\,Gbps, this will be further reduced to 12.5\,aJ/bit. Optically, most of the 17\,dBm optical power is dissipated at cryogenic due to large optical loss (50\,mW, 50\,pJ/bit), which is still comparable with cryogenic broadband electrical amplifiers (typical power consumption is 10-20\,mW). If assuming a combined 2.5\,dB/facet coupling loss and 0.05\,dB/cm propagation loss, a 1\,m-long SEOM is capable of optical readout of the RSFQ circuit at 10\,Gbps with 0\,dBm optical input (100\,fJ/bit, see Supplementary Information Sec.\,\rom{10}). For large-scale RSFQ readout, although each RSFQ high-frequency output requires one SEOM, it takes only one optical fiber to uplift the signal. For electrical readout, each output would require one coaxial cable, and each coaxial cable introduces tens of mW heat load from room temperate to 40\,K and a few mW heat load from 40\,K to 4\,K \cite{Krinner2019}.

Just as commercial EO modulators powering today's fiber networks, we believe SEOMs with significantly improved modulation efficiency and bandwidth can offer substantial advantages for future large-scale superconducting circuits. In this Article, we present the conceptual advances of SEOMs, including the development of meter-long modulators and a remarkable two orders of magnitude reduction in $V_{\rm{\pi}}$. Through careful device engineering, we achieve over 17\,GHz bandwidth on SEOM. Using this device, we have successfully demonstrated the direct optical readout of an RSFQ circuit for the first time. By further enhancing our fabrication processes and refining material processing to fully unlock the potential of the thin-film LN-superconductor platform, we anticipate several orders of magnitude further improvement in the modulation efficiency of meter-long SEOMs. This will enable the realization of scalable, low-power-consumption, and high-speed fully photonic links for superconducting circuits.

\section*{Acknowledgments}
\noindent This project was funded by IARPA’s SuperCables program through the ARO grant W911-NF-19-2-0115, and DOE Office of Science, National Quantum Information Science Research Centers, Co-design Center for Quantum Advantage (C2QA), Contract No. DE-SC0012704. We thank the Office of Naval Research for providing funding support in the construction of the RF interface through grant N00014-20-1-2134. Y.Z. acknowledges the support from Yale Quantum Institute. We extend our gratitude to Dr. Brad Liu for his assistance with RSFQ circuit module installation and operation. We would also like to acknowledge HYPRES for their contributions to the RSFQ circuit design and for granting permission to use the micrograph of the RSFQ circuit chip in this article. Special thanks go to Drs. Dmitri Kirichenko and Deep Gupta for their insightful discussions. We are grateful to Drs. Michael Gehl, Ben Palmer, Saewoo Nam, Deborah Van Vecheten, William Mayer, Willam Harrod for providing valuable technical and administrative support throughout this project. Finally, we thank Y. Sun,  S. Rinehart, L. McCabe, K. Woods and M. Rooks for assistance in the device fabrication.

\section*{Author contributions}
\noindent H.X.T. and  M.S. conceived the idea and experiment. M.S. fabricated the device and performed the experiment. J.X. and Y.X. helped with fabrication and experiments. S.W., R.C., W.F. and Y.Z. helped with device packaging and instrumentation. M.S. wrote the manuscript, and all authors contributed to the manuscript. H.X.T. supervised the work.

\section*{Competing interest}
\noindent The authors declare no competing interests.

\section*{Methods}
\noindent \textbf{Device fabrication.} The SEOM device is fabricated on $x$-cut LNOI wafer (NanoLN) featuring 600\,nm thin-film LN and 4.7\,$\mathrm{\micro m}$ buried oxide on a 500\,$\mathrm{\micro m}$-thick high-resistivity silicon substrate. The optical waveguides are defined using electron beam lithography (EBL) with hydrogen silsesquioxane (HSQ) as resist and etched by argon reactive ion etching. The optical waveguide is partially etched by 350\,nm with remaining 250\,nm slab. After stripping residual HSQ resist, 900\,nm HSQ is spun on chip and then defined by EBL as cladding layer. The chip is thermally annealed at 400\,$^\circ$C for 1 hour after resist development to turn the HSQ resist into silicon dioxide. The niobium electrodes (200-300\,nm) are defined using EBL with polymethyl methacrylate (PMMA) resist through liftoff process. The niobium film is deposited through electron beam evaporation in a ultra-high vacuum chamber ($2\times10^{-8}$\,torr during the deposition) at 10\,\r{A}/s deposition rate. The fabrication process flow is included in the Supplementary Information Sec.\,\rom{1}.

\noindent \textbf{Device characterization.} Cryogenic characterization of the EO bandwidth of the SEOM devices is performed in a closed-cycle low-vibration cryostat from Montana Instruments. The die-level chip is mounted on a 3-axis stack of micropositioners (Attocube). Through active alignment, photonic coupling to the SEOM device is through fiber array to on-chip optical grating couplers and electric connection is made using a multi-channel RF probe contacting on-chip superconductor electrode pads. Details regarding the calibration, data process and fitting of the EO response measurement is included in the Supplementary Information Sec.\,\rom{6}. Optical readout of the RSFQ circuit is performed using a packaged SEOM device in another closed-cycle custom-made cryostat\cmt{ made by Four Nine Design}. The packaged SEOM device and the RSFQ module are mounted on the front and back side of the 4\,K sample plate respectively. Details of the device packaging are included in the Supplementary Information Sec.\,\rom{8}.

\section*{Data availability}
\noindent The data that support the findings of this study are available from the corresponding authors upon reasonable request.

\section*{Code availability}
\noindent All relevant computer codes supporting this study are available from the corresponding author upon reasonable request.

\def\bibsection{\section*{References}}
\bibliography{new}

\end{document}


\title{Supplementary Information}
\maketitle
\author{} 
\onecolumngrid

\tableofcontents

\section{SEOM device fabrication} \label{sc:fab}

\begin{figure}[htbp]
\centering
\includegraphics[trim={0cm 0cm 0cm 0cm},clip]{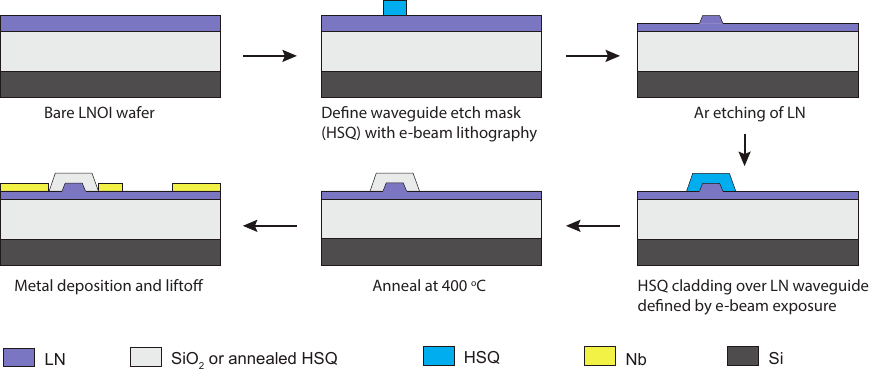}
\caption{SEOM fabrication process flow. 
}
\label{SFig:fab_flow}
\end{figure}

The fabrication process flow for SEOM is depicted in Fig.\,\ref{SFig:fab_flow}. It begins with a chip cleaved from an $x$-cut LNOI wafer (NanoLN), which has a 600\,nm-thick TFLN and 4.7\,$\mathrm{\mu m}$ buried oxide on 500\,$\mathrm{\mu m}$ high-resistivity silicon substrate. Firstly, a hydrogen silsesquioxane (HSQ) resist layer is spun on the chip to define optical waveguide using electron beam lithography (EBL). Next, the LN film is etched by 350\,nm using argon-base reactive ion etching. After stripping the residual resist, a second layer of 900\,nm-thick HSQ is spin-coated on the chip, followed by EBL to form HSQ cladding. In this step, the dose of HSQ is gradually reduced at the edge to give the HSQ cladding structure a sloped sidewall after development. The chip is then annealed at 400\,$^\circ$C for 1\,h. HSQ is chosen for LN waveguide cladding instead of PECVD deposition to suppress the photorefractive effect \cite{Xu2021a}. The Nb electrodes are defined through liftoff process using polymethyl methacrylate (PMMA) as resist. Nb is evaporated using electron beam in an ultra-high-vacuum chamber with $2\times 10^{-8}$\,torr pressure during the evaporation process. The deposition rate is approximately 10\,\r{A}/s.

\section{RSFQ driver circuits}

\begin{figure}[htbp]
\centering
\includegraphics[trim={0cm 0cm 0cm 0cm},clip]{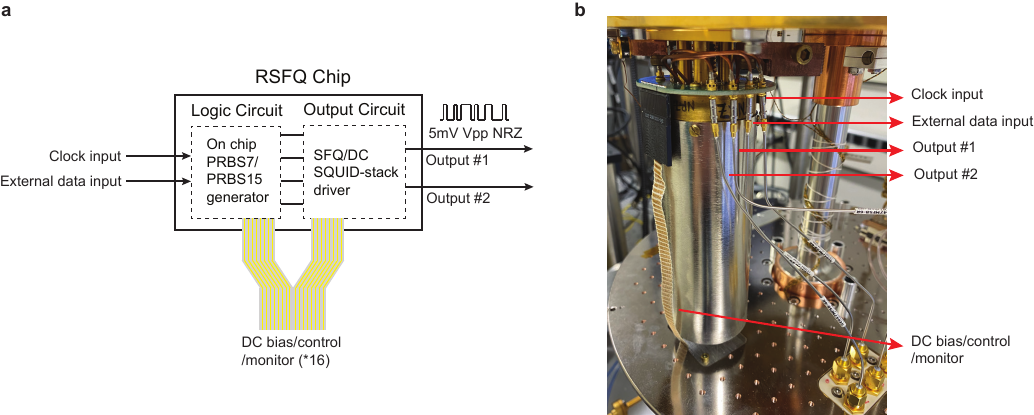}
\caption{\textbf{a}, Circuit diagram of the RSFQ NRZ signal generation. The logic circuit consists of PRBS7 and PRBS15 signal generator. The output circuit employs a SQUID-stack driver. Operation of the RSFQ circuit takes a clock input, external data input (optional) and 16 low frequency connection for bias, control and monitors. The circuit has two identical outputs. \textbf{b}, Picture of the packaged RSFQ stimulus module.}
\label{SFig:rsfq}
\end{figure}

The RSFQ chip is designed by HYPRES, Inc., and fabricated at MIT Lincoln Laboratory using SFQ5EE process \cite{Tolpygo2016}. It is capable of both NRZ and RZ format signal generation, although only NRZ signal output is utilized in our experiment. The circuit diagram NRZ signal generation is shown in Fig.\,\ref{SFig:rsfq}a. The operation of the RSFQ chip requires an external clock input and 16 low frequency connections for current bias, control and monitoring purposes. The logic circuit of the RSFQ chip has on-chip PRBS7 and PRBS15 signal generator, as described in \cite{Lehmann2017}. Additionally, there is an external data input configured to accept customized data generation. 

In SFQ logic, information is stored in magnetic flux quanta and transferred through voltage pulses. To interface with CMOS logic at room temperature, the output circuit employs SQUID-stack drivers \cite{Inamdar2009} called Ostrich \cite{Gupta2019}, which generate approximately 5\,mVpp digital signal output. Within the data driver, an integrated SFQ/DC converter utilizes a single SQUID driven by a flip-flop. When the flip-flop is in '1' state, it generates a train of SFQ pulses; while in the '0' state, no output is produced. This pulse train is directed to a stack of SQUIDs, and the output signals from the SQUID stack are summed to generate the final output. By carefully matching the excitation signal and the output signal from the SQUID stack, high speed data rate can be achieved. The RSFQ circuit we use is capable of data rates of up to 10\,Gpbs. Two identical data drivers are implemented in the RSFQ circuits, providing two identical outputs (output 1 and output 2 in Fig.\,\ref{SFig:rsfq}a). While the RSFQ chip can operate at higher data rates, the voltage waveform at the module's interfaces may degrade due to the limited interconnect bandwidth.

The RSFQ chip is enclosed in a cryogenic stimulus module, which incorporates a magnetic shield and provides all necessary thermal connectivity and electrical connections to the chip through pressure contact \cite{Dotsenko2017}. The RSFQ stimulus module interfaces with external equipment through two types of connections located at the base printed circuit board (PCB). The base PCB is visible near the top of the Fig.\,\ref{SFig:rsfq}b. Four high-speed signal connections (clock, data input and two outputs) utilize female G3PO coaxial connectors by Corning Gilbert. For the low frequency connections (current bias, monitoring and control signals), a Min-E-Con 16-pin connector is used, as shown in Fig.\,\ref{SFig:rsfq}b. The low-frequency wiring employs a copper loom, while coaxial cables are used for the high-speed signal. The complete stimulus module is mechanically mounted onto the back side of 4\,K plate of cryostat. Heaters and thermal sensors are mounted on the module to tune the working temperature of the RSFQ circuit for optimized signal quality.

\section{Quantum optics description of traveling-wave electro-optic modulation}
\label{sc:theory}

With their remarkably low $V_\pi$ values, SEOMs can operate as broadband microwave-to-optics transducers and offer competitive transduction efficiency compared to cavity-based electro-optic quantum transducers. Here we present a theoretical description of the traveling-wave electro-optic modulation process within the quantum optics framework. Our approach follows the quantum Hamiltonian treatment outlined in \cite{JESipe2016}. We use the theoretic results derived here to fit our experimental results. Note that here we focus more on the classical aspects of the electro-optic modulation process, and vacuum fluctuation and quantum noise terms are omitted.

\begin{figure}[htbp]
\centering
\includegraphics[trim={0cm 0cm 0cm 0cm},clip]{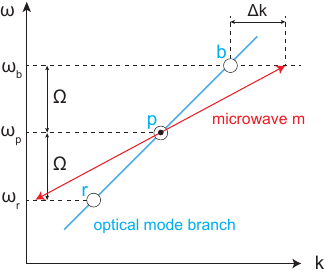}
\caption{Dispersion consideration in traveling-wave electro-optic modulators. A carrier (pump) mode $p$ at $\omega_p$ is scattered by a microwave mode $m$ at frequency $\Omega$ into the sidebands $b$ at $\omega_p + \Omega$ and $r$ at $\omega_p - \Omega$. The velocity difference between  microwave and optics leads to a wavenumber mismatch $\Delta k$.}
\label{SFig:dispersion}
\end{figure}

\subsection{Hamiltonian for traveling-wave electro-optic modulation}

Consider an optical waveguide coupled with a co-propagating microwave transmission line, the electrical and optic modal overlap leads to the electro-optic coupling. We assume both optical waveguide and microwave transmission line are invariant in the propagation direction ($z$ axis) and have the same length $L$. With periodic boundary condition, the wave vector of the microwave and optical modes must be an integer times $2\pi/L$. As $L\rightarrow\infty$, the mode spectrum becomes continuous and mode branches begin to form. Here we consider the dispersion of optical and microwave branches of traveling-wave electro-optics, as depicted in Fig.\,\ref{SFig:dispersion}. In this case, the phase matching requirement is straightforward: the slope of the optical mode dispersion curve (group velocity) must match that of the microwave (phase velocity), known as velocity matching in traveling-wave electro-optic modulator design. Note that if this phase matching requirement is satisfied over a wide frequency range, high-order optical sidebands can be generated in cascade. In the following analyses, we do not consider the higher sidebands due to their low conversion efficiency. Hence, we have four modes of interests in this model, optical pump $p$, blue and red sidebands $b$ and $r$, and microwave mode $m$. The system Hamiltonian can then be expressed as follows, 
\begin{equation}
    H_{sys}=H_p+H_b+H_r+H_m+H_{V},
    \label{eq:sysHamiltonian}
\end{equation}
where $H_p$, $H_b$, $H_r$ and $H_m$ are the Hamiltonian for the four modes of interests and $H_{V}$ is the electro-optic coupling term. In the following, we derive the expression for each component of the system Hamiltonian.

The electric field of the four modes shown in Fig.\,\ref{SFig:dispersion} can be expressed in terms of the continuous mode creation and annihilation operators \cite{Blow1990a},
\begin{eqnarray} \label{electricfield}
    \mathbf{E}_{l}&=&\int\frac{l(k)\mathbf{u}_{lk}(x,y)e^{ikz}+H.c.}{\sqrt{4\pi\int\epsilon_{i}u_{lki}u^*_{lki}dxdy/\hbar\omega_l(k)}}dk,
\end{eqnarray}
where $l(k)$ is the annihilation operators of the continuous optical or microwave electromagnetic modes of wavevector $k\mathbf{e}_z$, in which $l=p, b, r$ or $m$; the electric field mode profile of each mode branch at wave vector $k$: $\mathbf{u}_{lk}(x,y)=u_{l ki}(x,y)\mathbf{e}_i$, where $\mathbf{e}_i$ is the direction vector, $i=x, y$ or $z$ and Einstein summation convention is applied. The denominators in the equation are for normalization purpose and will disappear in the expression of Hamiltonian. The commutation relations of the continuous annihilation and creation operators indexed by $l$ and $l'$ reads below \cite{Blow1990a}
\begin{eqnarray}
[l(k),l'(k')]=0,\;[l(k),l'^\dagger(k')]=\delta_{ll'}\delta(k-k'). \label{eq:ccommutation}
\end{eqnarray}
With Eq.\,\ref{electricfield} and neglecting vacuum fluctuations, the Hamiltonian of each mode branch is
\begin{eqnarray}\label{eq:generalHamiltonian}
H_{l}&=&\int\epsilon_{i}E_{l i}E_{l i}^*dxdydz = \int\hbar\omega_{l}(k)l^{\dagger}(k)l(k)dk. \label{alphaHamil}
\end{eqnarray}
In Eq.\,\ref{alphaHamil}, the $\frac{1}{2}$ factor of the electric field energy density is multiplied by 2 to account for the magnetic field energy as the electric field and the magnetic field have the same amount of energy.

With the continuous mode operators, a wave packet can be mathematically expressed by the envelop operator, which is defined as
\begin{eqnarray}
    \widetilde{L}_{\omega}(z) &=& \frac{1}{\sqrt{2\pi}}\int l_ke^{i(k-k_l(\omega))z}dk,
    \label{eq:envelopDef}
\end{eqnarray}
where $\widetilde{L}_{\omega}(z)$ is the envelop of the wave packet with central frequency $\omega$, $L=P, B, R$ or $M$ in our model. Through inverse Fourier transformation, we have
\begin{eqnarray}
\label{inversetransform}
    l_k=\frac{1}{\sqrt{2\pi}}\int \widetilde{L}_{\omega}(z)e^{-i(k-k_{l}(\omega))z}dz.
\end{eqnarray}

The commutation relation of the envelop operators can be derived from the commutation relation of the continuous creation and annihilation operators in Eq.\,\ref{eq:ccommutation}, which gives
\begin{subequations}\label{eq:ecommutation}
\begin{eqnarray}
[\widetilde{L}_{\omega}(z),\widetilde{L}_{\omega'}(z')] &=& 0,
\\
{[}\widetilde{L}_{\omega}(z),\widetilde{L}^\dagger_{\omega'}(z')] &=& e^{i(k_l(\omega')-k_l(\omega))z}\delta(z-z').
\end{eqnarray}
\end{subequations}

Now we express the Hamiltonian using the envelop operators. Substituting Eq.\,\ref{inversetransform} to Eq.\,\ref{eq:generalHamiltonian} and using the expansion expression for $\omega_{l}(k)=\sum_{n=0}^{\infty}\frac{1}{n!}\frac{\partial^{n} \omega_{l}}{\partial k^n}\mid_{\omega}(k-k(\omega))^{n}$ , the Hamiltonian for mode $l$ becomes
\begin{eqnarray}
H_{l}=\int\hbar \widetilde{L}^{\dagger}(z,\omega)(\sum_{n=0}^{\infty}\frac{1}{n!}\frac{\partial^{n} \omega_{l}}{\partial k^n}\mid_{\omega}(-i\frac{\partial}{\partial z})^{n}\widetilde{L}(z,\omega))dz.\label{envelophamil}
\end{eqnarray}
To arrive at the expression above we used equation
\begin{eqnarray}
\int k^{n}f(x)e^{ikx}dkdx&=&(-i)^{n}\int f(x)\frac{\partial^{n}e^{ikx}}{\partial x^{n}}dkdx\nonumber
\\
&=&(-i)^{n}\int\frac{\partial^{n}f(x)}{\partial x^{n}}e^{ikx}dkdx.
\end{eqnarray}

Note that in Eq.\,\ref{eq:envelopDef}, the integration is over all the wavenumber and physically it means that the envelop operator represents the photons in all the frequency range. In real experiments, what we encounter is narrow-band signals like the laser. In this narrow bandwidth wavepacket assumption, we only consider the first two terms of the expansion and the Hamiltonian can be simplified to
\begin{eqnarray}
H_l=\int(&&\hbar\omega\widetilde{L}^{\dagger}_{\omega}(z)\widetilde{L}_{\omega}(z)-i\hbar v_l\widetilde{L}^{\dagger}_{\omega}(z)\frac{\partial\widetilde{L}_{\omega}(z)}{\partial z})dz,
\end{eqnarray}
where $v_l=\frac{\partial\omega_l}{\partial k}\mid_{\omega}$ is the group velocity at frequency $\omega$ for mode branch $l$. The first term assumes that the signal is monochromatic and $\widetilde{L}^{\dagger}_{\omega}(z)\widetilde{L}_{\omega}(z)$ is the photon number density, while the second term is a correction to the monochromatic assumption, because it is a wavepacket instead.

Next, we find the expression for $H_V$ in terms of the envelop operators. If we take $p$ as the pump, there are two parts in $H_V$: parametric down conversion between $p$, $r$ and $m$ and coherent electro-optic conversion (up conversion) between $p$, $b$ and $m$. So we have
\begin{eqnarray}
H_{V}=\frac{1}{2}(\int r_{ijk}\epsilon_i\epsilon_jE_{pi}E_{bj}E_{mk}dxdydz + \int r_{ijk}\epsilon_i\epsilon_jE_{pi}E_{rj}E_{mk}dxdydz),
\end{eqnarray}
where $r_{ijk}$ is the electro-optic component and $(i,j,k)$ is the permutation of $(x,y,z)$. In Eq.\,\ref{electricfield} the electric field mode profile $u_{l k}(x,y)$ is a function of the wavelength, but within a narrow frequency range, it is a good assumption that the mode profile does not change. Given this, substitute Eq.\,\ref{electricfield} into $H_{V}$, we have
\begin{eqnarray}
H_{V}&=&\int (\frac{\hbar g_0e^{i\Delta k z}}{(2\pi)^{\frac{3}{2}}}p(k_1)b^\dagger(k_2)m(k_3)+H.c.)dk_1dk_2dk_3dz + \nonumber \\
& &\int (\frac{\hbar g^{'}_0e^{i\Delta k z}}{(2\pi)^{\frac{3}{2}}}r(k_1)p^\dagger(k_2)m(k_3)+H.c.)dk_1dk_2dk_3dz
\label{eq:Hint_original}
\end{eqnarray}

\begin{eqnarray}
\label{eq:vacuumcouplingrate}
\hbar g_{0}&=&\frac{1}{4\sqrt{2}}\frac{\int r_{ijk}\epsilon_{i}\epsilon_{j}u_{pi}u_{bj}^{*}u_{mk}dxdy}{\prod_{l=p,b,m}\sqrt{\int\epsilon_{l i}u_{l i}u_{l i}^{*}dxdy/\hbar\omega_{l}}}, \hbar g^{'}_{0}=\frac{1}{4\sqrt{2}}\frac{\int r_{ijk}\epsilon_{i}\epsilon_{j}u_{ri}u_{pj}^{*}u_{mk}dxdy}{\prod_{l=r,p,m}\sqrt{\int\epsilon_{l i}u_{l i}u_{l i}^{*}dxdy/\hbar\omega_{l}}}
\end{eqnarray}
where $\Delta k=k_1-k_2+k_3$ and $g_0$ is the vacuum electro-optic coupling strength for the traveling-wave case. Suppose that the central frequency of the wavepackets of mode $p$, $b$ , $r$ and $m$ are $\omega_p$, $\omega_b$, $\omega_r$ and $\Omega$ respectively, then Eq.\,\ref{eq:Hint_original} can be written in terms of envelop operators as
\begin{eqnarray}
H_{V}&=&\int(\hbar g_0\widetilde{P}(z,\omega_{p})\widetilde{B}^{\dagger}(z,\omega_{s})\widetilde{M}(z,\Omega)e^{i\Delta k z}+H.c.)dz + \nonumber
\\
&&\int(\hbar g^{'}_0\widetilde{R}(z,\omega_{p})\widetilde{P}^{\dagger}(z,\omega_{s})\widetilde{M}(z,\Omega)e^{i\Delta k' z}+H.c.)dz
\label{eq:Hint_appen}
\end{eqnarray}
where $\Delta k=k_p(\omega_p)-k_b(\omega_b)+k_c(\Omega)$ and $\Delta k'=k_r(\omega_r)-k_p(\omega_p)+k_c(\Omega)$ are the phase mismatch terms. Here we have derived the system Hamiltonian in terms of the envelop operators, which can be applied to general electro-optic interaction process. It will be simplified by the dispersion and phase matching conditions when applied to traveling-wave electro-optic modulation process, as discussed in the next section.

\subsection{Dispersion and phase matching in traveling-wave electro-optics}

With the optical pump at frequency $\omega_p$ and microwave signal at $\Omega$, the blue and red sideband frequencies are $\omega_b = \omega_p + \Omega$ and $\omega_r = \omega_p - \Omega$, as shown in Fig.\,\ref{SFig:dispersion}. In the traveling-wave electro-optic modulator scenario, there is only one optical mode branch. Because the microwave frequency is far below that of the optics, the optical dispersion can be simplified to a linear dispersion relation, thus, 
\begin{eqnarray}
\label{eq:wavevectorRelation}
k_b(\omega_p+\Omega) = k_p + v_o\Omega,
\\
k_r(\omega_p-\Omega) = k_p - v_o\Omega,
\end{eqnarray}
where $v_o$ is the optical group velocity at $\omega_p$. With this dispersion relation, the $\Delta k$ and $\Delta k'$ in Eq.\,\ref{eq:Hint_appen} become
\begin{eqnarray}
\label{eq:wavevectorMismatch}
\Delta k = \Delta k' = (v_{m,p} - v_o)\Omega,
\end{eqnarray}
where $v_{m,p}$ is the microwave phase velocity. Eq.\,\ref{eq:wavevectorMismatch} shows how the wavenumber mismatch is related with the velocity mismatch. To realized a high-bandwidth electro-optic modulation, $v_{m,p}$ needs to be a constant over that frequency bandwidth and matches to $v_o$. As $n_g = n_p + f\frac{\partial n_p}{\partial f}$, where $n_g$, $n_p$ and $f$ are the group index, phase index and frequency, we have $n_g=n_p$ when the microwave is dispersionless. So, the requirements for high bandwidth are that: the microwave is dispersionless and its phase velocity (or group velocity) matches to the group velocity of optics. 

Because the optical modes $p, b$ and $r$ are in the same optical mode branch and very close in frequency, the values $g_0$ and $g'_0$ in Eq.\,\ref{eq:vacuumcouplingrate} are the same.

\subsection{Dynamic equations and solutions} \label{section_dynamics}

We are now prepared to derive the system dynamic equations. Using the commutation relation Eq.\,\ref{eq:ecommutation}, we obtain the Heisenberg equation in terms of the envelop operators
\begin{subequations}
\label{eq:originalsysHeisenberg}
\begin{eqnarray}
\frac{\partial\widetilde{P}}{\partial t}&=&-(i\omega_{p}+\frac{\kappa_{o}}{2})\widetilde{P}-v_o\frac{\partial\widetilde{P}}{\partial z}-ig_{0}^{*}\widetilde{B}\widetilde{M}^{\dagger}e^{-i\Delta kz}-ig_{0}\widetilde{R}\widetilde{M}e^{i\Delta kz},
\\
\frac{\partial\widetilde{B}}{\partial t}&=&-(i\omega_{b}+\frac{\kappa_{o}}{2})\widetilde{B}-v_{o}\frac{\partial\widetilde{B}}{\partial z}-ig_{0}\widetilde{P}\widetilde{M}e^{i\Delta kz},
\\
\frac{\partial\widetilde{R}}{\partial t}&=&-(i\omega_{r}+\frac{\kappa_{o}}{2})\widetilde{R}-v_{o}\frac{\partial\widetilde{R}}{\partial z}-ig_{0}^{*}\widetilde{P}\widetilde{M}^{\dagger}e^{-i\Delta kz},
\\
\frac{\partial\widetilde{M}}{\partial t}&=&-(i\Omega+\frac{\kappa_{m}}{2})\widetilde{M}-v_{m}\frac{\partial\widetilde{M}}{\partial z}-ig_{0}^{*}\widetilde{P}^{\dagger}\widetilde{B}e^{-i\Delta kz}-ig_{0}^{*}\widetilde{R}^{\dagger}\widetilde{P}e^{-i\Delta kz}.
\end{eqnarray}
\end{subequations}
where $\kappa_o$, $\kappa_m$ and $v_o$, $v_m$ are the intrinsic loss rates and group velocities of the optical modes in the optical waveguide and microwave mode in the transmission line respectively. We remove the fast oscillating component in $\widetilde{P}$, $\widetilde{B}$, $\widetilde{R}$ and $\widetilde{M}$ by introducing $P=\widetilde{P}e^{i\omega_{p}t}$, $B=\widetilde{B}e^{i\omega_b t}$, $R=\widetilde{R}e^{i\omega_r t}$ and $M=\widetilde{M}e^{i\Omega t}$. During the practical device operation, the pump is a constant laser input with high intensity to boost the electro-optic conversion efficiency. Because of its much stronger intensity than that of the signal light and microwave, the back-action of electro-optic interaction on the pump is negligible and the pump can be taken as a classical light source. Therefore in the steady-state,
\begin{eqnarray}
P(z,t)=P_0 e^{-\frac{\kappa_o}{2v_o}z}
\end{eqnarray}
where $P_0$ is a constant and the exponential decay term indicates the dissipation in the optical waveguide of mode $p$. Suppose that at $z=0$, the pump power is denoted as $P_{opt}=v_o\hbar\omega_{p}|P_0|^2$. If the phase of the pump is set to be $0$,
\begin{eqnarray}
P_0=\sqrt{\frac{P_{opt}}{v_o\hbar\omega_{p}}}.
\end{eqnarray}
Combining this equation and Eq.\,\ref{eq:originalsysHeisenberg}, we can derive the temporal-spatial dynamic equations of the system
\begin{subequations}
\label{eq:LFsysHeisenberg}
\begin{eqnarray}
\frac{\partial B}{\partial t}+v_{o}\frac{\partial B}{\partial z}&=&-\frac{\kappa_{o}}{2}B-ig_{0}P_0 e^{(-\frac{\kappa_o}{2v_o}z + i\Delta k)z} M,
\\
\frac{\partial R}{\partial t}+v_{o}\frac{\partial R}{\partial z}&=&-\frac{\kappa_{o}}{2}R-ig_{0}^{*}P_0 e^{(-\frac{\kappa_o}{2v_o}z-i\Delta k)z} M^{\dagger},
\\
\frac{\partial M}{\partial t}+v_{m}\frac{\partial M}{\partial z}&=&-\frac{\kappa_{m}}{2}M-ig_{0}^{*} P_0^* e^{(-\frac{\kappa_o}{2v_o}z-i\Delta k)z} B - ig_{0}^{*} P_0 e^{(-\frac{\kappa_o}{2v_o}z-i\Delta k)z} R^{\dagger}.
\end{eqnarray}
\end{subequations}

With these dynamic equations, we solve for the intensity of the converted optical sidebands and the microwave mode with microwave input $M(z) = M_0$ at $z=0$. For the steady-state solution, where the time differential terms in Eq.\,\ref{eq:LFsysHeisenberg} are zero, the solution to Eq.\,\ref{eq:LFsysHeisenberg} is
\begin{subequations}
\begin{eqnarray}
    B(z) &=& -i\frac{g_0 P_0}{v_o} M_0 e^{-(\alpha_o/2) z} \frac{1 - e^{-(\alpha_m/2 - i\Delta k)z}}{\alpha_m/2 - i\Delta k},
    \label{eq:solution_B}
    \\
    R(z) &=& -i\frac{g_0 P_0}{v_o} M_0 e^{-(\alpha_o/2) z} \frac{1 - e^{-(\alpha_m/2 + i\Delta k)z}}{\alpha_m/2 + i\Delta k},
    \label{eq:solution_R}
    \\
    M(z) &=& M_0 e^{-(\alpha_m/2) z}.
    \label{eq:solution_M}
\end{eqnarray} \label{eq:solution}
\end{subequations}

\noindent Here $\alpha_o = \frac{\kappa_o}{v_o}$ and $\alpha_m = \frac{\kappa_m}{v_m}$ are the optical and microwave power propagation loss. From this result we find that the microwave intensity is not affected by the electro-optic interaction. This is because: in the parametric down conversion process, one microwave photon is generated with one red sideband optical photon; in the coherent electro-optic up conversion process, one microwave is absorbed together with a pump photon to generate a blue sideband optical photon. While for the traveling-wave electro-optic modulation process, where the parametric down conversion and up conversion happen simultaneously, the net effect on the microwave signal is that its photon number remain the same if there is no propagation loss.

Under the condition that the modulation length is short (in the conventional modulators' case), $|(\alpha_m/2 \pm i\Delta k)z| \ll 1$, the last terms in solutions \ref{eq:solution_B} and \ref{eq:solution_R} reduce to $z$. The out-of-chip microwave-to-optical transduction efficiency is then
\begin{eqnarray}
    \eta = \frac{v_o |B(z)|^2}{v_m M_0^2} = \frac{(g_0 P_0)^2}{v_o v_m} e^{-\alpha_o z} z^2.
    \label{eq:efficiency}
\end{eqnarray}
As $V_\pi$ is inversely proportional to modulation length, it implies that the efficiency is inversely proportional to the square of $V_\pi$. In the next section we point out the connection between $V_\pi$ and the vacuum electro-optic coupling strength $g_0$, and Eq.\,\ref{eq:efficiency} will be the same as the result derived in \cite{Youssefi2020}.

When the modulation length is long, optical and microwave propagation loss and the velocity mismatch $\Delta k$ come into play. In the last term of \ref{eq:solution_B} and \ref{eq:solution_R}, if we ignore the propagation loss term and keep only the phase mismatch, it becomes the sinc function, suggesting that if the velocity mismatch is significant enough, there is null frequencies in the electro-optic response. This will be illustrated in Section\,\ref{sc:EOresponse} where we use these equations to fit our experimental results. If there is no velocity mismatch, the microwave and optical loss become the fundamental limit on the conversion efficiency. Using the results obtained here, we project in Section \ref{sc:projection} the ultimate achievable performance of the SEOM devices.

\subsection{Connection between $g_0$ and $V_\pi L$}

In this section we show how the traveling-wave coupling strength $g_0$ used in the quantum optics language is related to $V_\pi L$, the conventional figure of merit to characterize the efficiency of electro-optic modulators. Suppose that a voltage $V$ is applied across the electrodes, and the induced optical mode index change is $\delta n (V)$, we have
\begin{eqnarray}
    \frac{\delta n (V)}{n} = \frac{1}{2}\frac{\int r_{ijk}\epsilon_{i}\epsilon_{j}u_{oi}u_{oj}^{*}u_{mk}(V)dxdy}{\int\epsilon_{i}u_{o i}u_{o i}^{*}dxdy/\hbar\omega}
    \label{eq:dn}
\end{eqnarray}
where $\mathbf{u}_{o}(x,y)=u_{oi}(x,y)\mathbf{e}_i$, is the optical mode profile, $i=x, y$ or $z$. $\mathbf{u}_{m}(V)(x,y)=u_{mi}(V)(x,y)\mathbf{e}_i$ is the microwave mode profile when voltage $V$ is applied. Here we see a similarity between this equation and the definition of $g_0$ in Eq.\,\ref{eq:vacuumcouplingrate}. Upon inserting $\frac{\delta n (V)}{n}$ into Eq.\,\ref{eq:vacuumcouplingrate}, we have
\begin{eqnarray}
    g_0 &=& \frac{\omega}{2\sqrt{2}}\frac{\delta n(V)}{n} \sqrt{\frac{\hbar \Omega}{\int\epsilon_{i}u_{m i}(V)u_{m i}^{*}(V)dxdy}} \nonumber
    \\
    & = & \frac{\sqrt{\hbar \Omega v_m} v_o }{2}\frac{\omega \delta n(V) z}{v_o n} \frac{1}{z\sqrt{2 v_m \int\epsilon_{i}u_{m i}(V)u_{m i}^{*}(V)dxdy}} \nonumber
    \\
    & = & \frac{\sqrt{\hbar \Omega v_m} v_o }{2} (k \delta n(V) z) \frac{1}{z\sqrt{P_m(V)}}
    \label{eq:g0anddn}
\end{eqnarray}
where $k$ is the optical wavenumber in vacuum and $k \delta n(V) z$ is the optical phase change accumulated after an interaction length of $z$; $P_m(v)$ is the microwave input power with voltage amplitude $V$. For a given $z$, when $V = V_\pi$, $k \delta n(V_\pi) z = \pi$ and $P_m(V_\pi) = V_\pi^2/(2Z_0)$, $Z_0$ is the transmission line impedance. Therefore we have
\begin{eqnarray}
    g_0 &=& \frac{\pi v_o \sqrt{\hbar\Omega v_m Z_0}}{\sqrt{2}} \frac{1}{V_\pi L}
    \label{eq:g0andVpiL}
\end{eqnarray}
Substituting Eq.\,\ref{eq:g0andVpiL} into Eq.\,\ref{eq:efficiency}, we can express the conversion efficiency in terms of $V_\pi$,
\begin{eqnarray}
    \eta &=& P_{opt}\frac{\pi^2}{2} \frac{\Omega}{\omega} \frac{Z_0}{V_\pi^2}.
    \label{eq:efficiency_dn}
\end{eqnarray}

\begin{figure}[htbp]
\centering
\includegraphics[trim={0cm 0cm 0cm 0cm},clip]{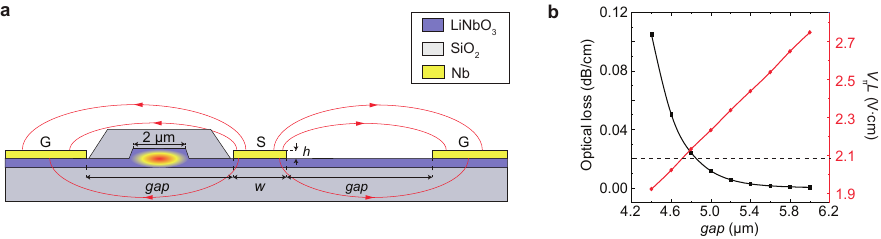}
\caption{ Numerical simulation of metal induced optical loss and $V_\pi L$.}
\label{SFig:VpiL}
\end{figure}

The $V_\pi L$ value is primarily determined by the gap between the modulation electrodes, and the smallest possible gap distance is constrained by metal induced optical loss when the electrodes are in close proximity to the waveguide, as illustrated in the simulation shown in Fig.\,\ref{SFig:VpiL}b. In order to ensure that the metal loss is significantly lower than the intrinsic loss of our fabricated optical waveguide (0.2\,dB/cm), we aim for a metal loss of 0.02\,dB/cm. Consequently, the value of  $V_\pi L$ exceeds 2.1\,V$\cdot$cm. In our real device design, we choose the gap to be 5.2\,$\mathrm{\mu m}$.

\section{Optical transmission of SEOM}

\begin{figure}[htbp]
\centering
\includegraphics[trim={0cm 0cm 0cm 0cm},clip]{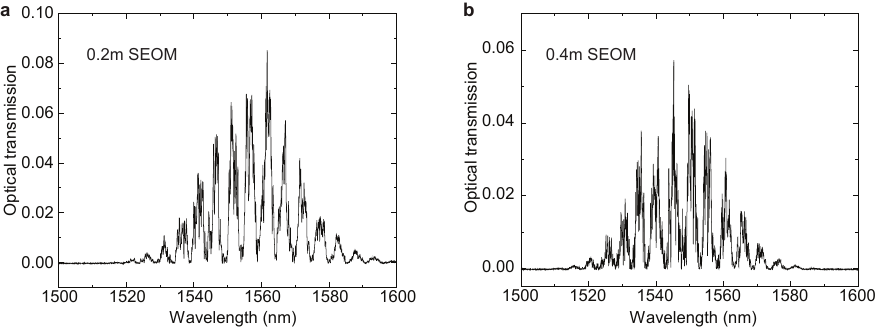}
\caption{ Optical transmission of two SEOM devices with total length of 0.2\,m and 0.4\,m. Optical grating couplers are employed for optical coupling into the device. The oscillation in the spectrum is due to the interference of two unbalanced optical arms. }
\label{SFig:optical_trans}
\end{figure}

Optical transmission spectrum of two SEOM devices are shown in Fig\,\ref{SFig:optical_trans}. The two SEOM devices have total modulation length of 0.2\,m and 0.4\,m respectively. Optical grating couplers are used for coupling from optical fibers to the on-chip SEOM devices. The central wavelength of the grating coupler is designed to be around 1550\,nm. The oscillation on the spectrum is due to the interference of the two unbalanced optical arms of the SEOM device. We intentionally design MZI modulator to be unbalanced such that the working point can be tuned by the optical wavelength. After the optical grating coupler, a short section of 0.8\,$\mathrm{\mu m}$-wide single-mode waveguide is used to filter out high-order modes. This narrow waveguide is then tapered to 2\,$\mathrm{\mu m}$-wide wavegude in the device area to reduce the optical propagation loss. Due to the extremely long optical waveguide length, the optical transmission spectrum is not as clean as short waveguides. We attribute this to the waveguide sidewall roughness/fabrication imperfections induced imbalance between the two optical arms, reflections and conversion into high-order modes.

\section{ {$V_\pi$} change from room temperature to cryogenic temperature } 

\begin{figure}[htbp]
\centering
\includegraphics[trim={0cm 0cm 0cm 0cm},clip]{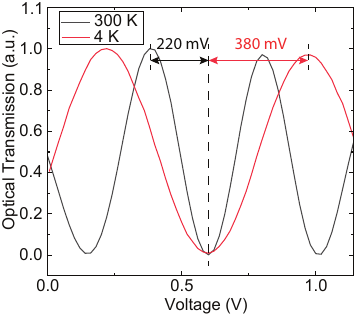}
\caption{ $V_\pi$ measurement at room temperature and 4\,K. A 0.2\,m-long SEOM is measured at both room temperature and 4\,K. Its $V_\pi$ increases from 220\,mV to 380\,mV. }
\label{SFig:RT_CT_Vpi}
\end{figure}

Cryogenic temperature dependence of the electro-optic coefficient of lithium niobate has been studied \cite{Herzog2008a}, and it shows that the $r_{33}$ component (which is the component utilized in most electro-optic modulators) decreases by 20\% at 7\,K. We also observe a $V_\pi$ increases by 70\% from room temperature to cryogenic temperature in our SEOM. For a 0.2\,m SEOM, its $V_\pi$ at room temperature is 220\,mV, and 380\,mV at 4\,K, as shown in Fig\,\ref{SFig:RT_CT_Vpi}. Compared with other studies, in Ti-diffusion waveguide electro-optic modulators, $V_\pi$ decrease of 10\% \cite{Youssefi2020} and increase of 74\% \cite{Thiele2022} are reported. On a ridge waveguide based electro-optic modulator on LNOI, a $V_\pi$ increase of 15\% is reported \cite{Lomonte2021b}. In all the modulators mentioned above, $r_{33}$ component is commonly utilized in the modulator design. We think the variance on the reported values might be due to different material growth methods and fabrication conditions. With some of the reported results showing that the $V_{\rm{\pi}}$ decreases or does not substantially change at cryogenic temperature, it is possible that the 70\% $V_{\rm{\pi}}$ increase could be reduced or avoided, but this is subject to further investigations.

\section{Electro-optic bandwidth of SEOM} \label{sc:EOresponse}

In the main text we present the measured electro-optic response along with fitting. Here we provide the detailed data analysis and fitting procedure. We use theoretical results Eq\,\ref{eq:solution} obtained in section \ref{section_dynamics} to model the electro-optic frequency responses, which is captured by the last term of Eq\,\ref{eq:solution_B} and \ref{eq:solution_R}. The parameters that depend on microwave frequency are $\alpha_m$, the microwave loss and $\Delta k$, the wavenumber mismatch. We will discuss how we determine the microwave loss and dispersion first and then use these parameters to model the experimental data.

\subsection{Microwave dispersion and loss}

\begin{figure}[htbp]
\centering
\includegraphics[trim={0cm 0cm 0cm 0cm},clip]{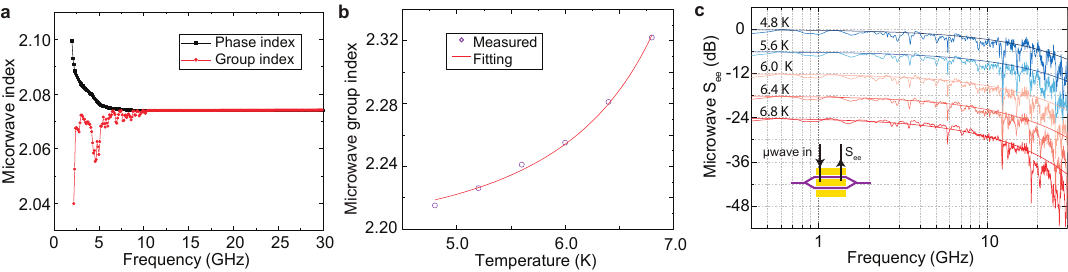}
\caption{\textbf{a}, Numerical simulation of the microwave phase and group indices. The cross-section of the microwave transmission line is as shown in Fig\,\ref{SFig:VpiL}a. The effective index (phase index) is directly simulated and the group index is derived from the phase index through the equation $n_g = n_p + f\frac{\partial n_p}{\partial f}$. Above 6\,GHz, the phase and group index are close to each other. \textbf{b}, Measured microwave group index of SEOM microwave transmission line at different temperatures with fitting. The temperature dependence of the kinetic inductance leads to the thermal tuning of group index. \textbf{c}, S$_{21}$ measurement of SEOM microwave transmission line at different temperatures. Solid lines are fitted envelop of the S$_{21}$ response.}
\label{SFig:uwave}
\end{figure}

We first investigate the microwave dispersion through simulation. As shown in Fig.\,\ref{SFig:uwave}a, the on-chip microwave transmission line mode is more dispersive at low frequency than at high frequency. Above 10\,GHz, the group and phase index of microwave are close to each other, indicating that the microwave mode is dispersionless in this frequency range. Exploiting this observation, we approximately use the group index of high frequency to represent the microwave phase index over the whole spectrum. The low frequency deviation is acceptable because when the frequency $\Omega$ is small,  $\Delta k = (v_{m,p} - v_o)\Omega$ is more tolerant to the velocity mismatch.

The group index of the microwave transmission line is determined experimentally using a VNA equiped with a time-domain option. By measuring the time it takes a microwave pulse to travel through the transmission line, we can calculate the microwave group velocity. The measured data is shown in Fig\,\ref{SFig:uwave}b. Fig.\,\ref{SFig:uwave}a illustrates the simulated phase and group velocities without considering  the effect of kinetic inductance. When the kinetic inductance is taken into account, the microwave group velocity is slowed down. The kinetic inductance is dependent on temperature, and so is the microwave group index. This temperature dependence is fitted by $y = a\sqrt{1 + b/\sqrt{1 - (T/T_c)^4}}$ \cite{tinkham2004introduction}, where $T$ is the temperature, $T_c$ is the superconductive transition temperature 8.0\,K (measured), and $a, b$ are the parameters to be fitted. The kinetic inductance provides us with the ability to tune the microwave velocity by adjusting the temperature, thereby aiding in velocity matching.

The microwave propagation loss is quantified by directly measuring the S21 spectrum, as shown in Fig\,\ref{SFig:uwave}c. To extract the frequency dependent propagation loss, a simple linear fitting is employed to determine the slope of the envelope of the transmission line's S21 response. In the linear fitting, the frequency is in linear scale and the S21 response is in dB (logarithmic) scale. In summary, we approximate  the microwave phase index using its group index, which has been validated through simulation. Both the microwave group index and propagation loss are determined experimentally.

\subsection{Electro-optic response}
After extracting microwave group index and frequency-dependent microwave loss from S$_{21}$ response in Fig.\,\ref{SFig:uwave}, we next fit the electro-optic response shown in Fig\,3d in the main text. The only unknown parameter is the optical group index that appears in $\Delta k$ as shown in Eq.\,\ref{eq:wavevectorMismatch}. In principle, we can use the simulated optical group index. However, the material's properties like the refractive index at cryogenic temperature is not well established from literature. The optical group index is thus treated as a fitting parameter inferred from the electro-optic response. In the scenario that the optical and microwave index are very mismatched, there is null frequency in the electro-optic response as predicted by Eq.\,\ref{eq:solution}. This null frequency in electro-optic response provides a convenient signature to determine the optical index and validate the model.

\begin{figure}[htbp]
\centering
\includegraphics[trim={0cm 0cm 0cm 0cm},clip]{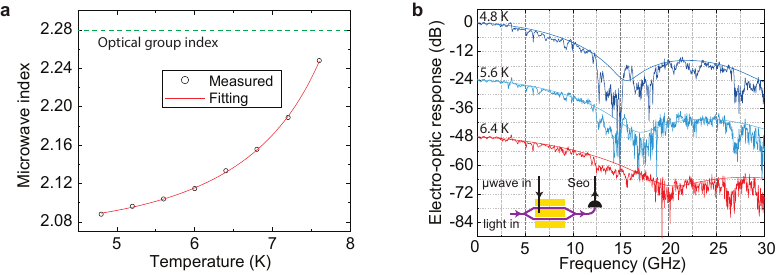}
\caption{\textbf{a}, Measured microwave group index. In this device, the microwave transmission line is intentionally designed to have a significantly lower group index than the optical group index of 2.28. The optical group index is obtained from the the fitting shown in the panel b. \textbf{b} Measured electro-optic response of this SEOM device at 4.8\,K, 5.6\,K and 6.4\,K, respectively. The response is shifted by 24\,dB for better visibility. Due to the index mismatch, the 3\,dB bandwidth of this device is only a few GHz and there is electro-optic response dip at 15\,GHz, 17\,GHz and 20\,GHz. The envelop of the curves are fitted with $n_o = 2.28$.}
\label{SFig:EO}
\end{figure}

For this purpose, we fabricate and measure another SEOM device with the same photonic waveguide structure as depicted Fig.\,3d, but with a slightly different transmission line geometry to create an intentional optical and microwave index mismatch. The measured microwave index is shown in Fig.\,\ref{SFig:EO}a, where a large index mismatch is shown. The electro-optic response of this device is shown in Fig\,\ref{SFig:EO}b, and the bandwidth of the responses is only a few GHz. Notably, the null frequencies appear at around 15, 17 and 20\,GHz. By fitting these electro-optic responses, we find that the best fit is achieved when the optical group index $n_o$ is set to 2.28, while the simulated optical group index using room temperature material coefficients is 2.25. Employing this optical group index, along with the measured microwave index and loss shown in Fig.\,\ref{SFig:uwave}, we obtain a good fit to the Fig.\,3d in the main text.

\section{ Eye diagram SNR analysis at low drive voltages } \label{sc:snr}

In this section we study the fundamental limiting factors on the signal-to-noise ratio (SNR) of eye diagrams in the low drive voltage region, where the peak-to-peak voltage ($V_{\rm{pp}}$) is much smaller than the half-wave voltage ($V_\pi$). The shot-noise-limited photoreceiver sensitivity has been widely studied, and different receiver architectures and modulation formats are employed to approach this limit \cite{HENRY1989, Olsson1989, Caplan2006, Lavery2011}. In this section we only focus on the direct detection with on-off keying. One distinction here is that instead of employing full-swing modulation ($V_{\rm{pp}} = V_\pi$), we utilize a small drive voltage. We will examine how the shot noise limit is translated to the optical carrier power, $V_{\pi}$ and working bias position in this specific scenario.

\subsection{Theoretical model}

\begin{figure}[htbp]
\centering
\includegraphics[trim={0cm 0cm 0cm 0cm},clip]{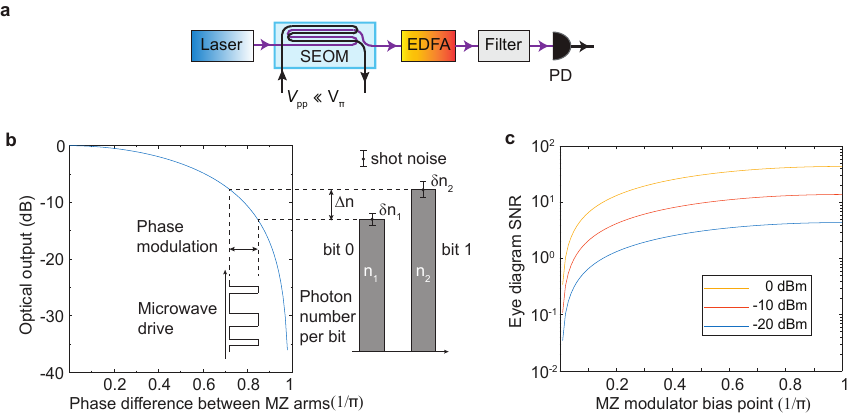}
\caption{ \textbf{a}, Eye diagram measurement setup. EDFA: Erbium-Doped Fiber Amplifier. PD: photodetector. \textbf{b}, Shot noise in a direct detection of on-off keying with small driving voltage. The transmission of the MZI modulator is determined by the phase difference of the two interference arms. The transmission bias point and the modulation induced phase change determine the photon number in bit 0 and 1. The two gray bars represent the photon number in the digital 0 and 1 bits and the error bars represent the shot noise. The photon number difference is $\Delta n = n_2 - n_1$, shot noise $\delta n_1 = \sqrt{n_1}$ and $\delta n_2 = \sqrt{n_2}$. SNR = $\Delta n / (\delta n_1 + \delta n_2)$. \textbf{c}, Eye diagram SNR limit. In this calculation $V_{\rm{pp}}/V_\pi = 0.01$ at 1\,Gbps. }
\label{SFig:SNR}
\end{figure}

Fig.\,\ref{SFig:SNR}a illustrate the eye diagram measurement setup. The MZI type SEOM is biased near the transmission null point to enhance the measurement SNR, which will be discussed further. As the transmission at this bias point is low, an erbium-doped fiber amplifier (EDFA) is employed to overcome the electrical noise in the photodetector and detection electronics. Here we assume a 100\,\% quantum efficiency for the PD. Additionally, an optical filter is placed between the EDFA and the PD to filter out the broadband amplified spontaneous emission noise. After optical amplification, the optical power swing encompasses a substantial portion of the photodetector's output range, thus overwhelming its thermal noise and dark current. Under this configuration, the dominant noise source is from the optical shot noise, as shown in Fig\,\ref{SFig:SNR}b. The optical output of an MZI type modulator is determined by phase difference between the two interference arms. When biased at a specific point, the phase modulation induces an optical output power change, encoding as digital bit 0 and 1. The photon number in each bit is $n_1$ and $n_2$, determined by the optical power and data rate. The shot noise in these two bits can be expressed as $\delta n_1 = \sqrt{n_1}$ and $\delta n_2 = \sqrt{n_2}$. We define the SNR\,$= (n_2-n_1)/ (\delta n_1 + \delta n_2)$.

The eye diagram SNR at 1Gbps, with $V_{\rm{pp}}/V_\pi = 0.01$, is plotted versus different bias points with different optical output powers in Fig.\,\ref{SFig:SNR}c. The optical power refers to the output power when the MZI is biased at its peak transmission. Numerical simulation demonstrates that the highest SNR is achieved when the MZI is biased near the transmission dip, owing to the steep slope and consequent high modulation sensitivity. However, the SNR improvement due to high extinction saturates as shot noise increases with decreasing transmission. Experimentally, the SEOM is typically biased between -15 to -20\,dB transmission. In our RSFQ circuit readout experiment, the SEOM's output power is approximately -3\,dBm, roughly corresponding to the yellow curve in Fig.\,\ref{SFig:SNR}c. Although the theoretical limit shows a SNR of about 40, out experimentally result is about 10\,dB away from this limit. Imperfections come from noise figure of the EDFA, wide filtration band of the optical filter and not deal quantum efficiency of the photodetector. SNR improvements can be made from the device parameters point of view. If bit 0 is biased at the null optical output, then $n_1 = 0$ and $\Delta n = n_2 \propto (V_{\rm{pp}}/V_\pi)^2 I_{o}/f$, where $I_o$ is the optical output power and $f$ is the data rate. So, the SNR is 
\begin{eqnarray}
    \mathrm{SNR} \propto \sqrt{\frac{I_0}{f}} \frac{V_{\rm{pp}}}{V_\pi}.
    \label{eq:snr}
\end{eqnarray}
The requirement for high SNR at a specific data rate speed aligns with the need for high out-of-chip transduction efficiency shown in Eq.\,\ref{eq:efficiency_dn}, which is quadratic in $V_{\pi}$ and linear in optical power. With a fixed optical input power, the output $I_o$ is determined by the propagation loss and modulation length, while the $V_{\pi}$ is inversely proportional to the modulation length. This tradeoff is captured by Eq.\,\ref{eq:efficiency}. In the following section, we will discuss strategies to enhance the SNR (or equivalently electro-optic transduction efficiency).

\subsection{Experimental results}

\begin{figure}[htbp]
\centering
\includegraphics[trim={0cm 0cm 0cm 0cm},clip]{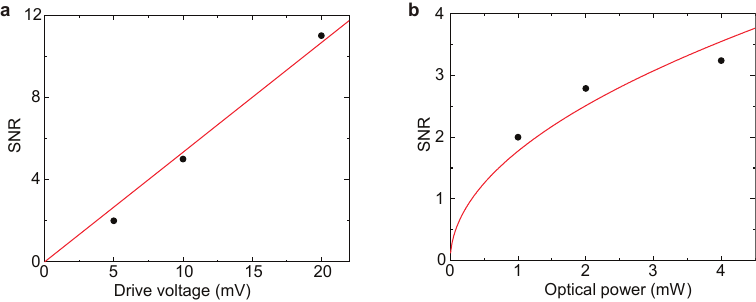}
\caption{ 
\textbf{a}, Eye diagram SNR dependence with drive voltage. The eye diagrams are taken with 17\,dBm optical input power at 1\,Gbps. The red curve is linear fitting of the experimental data. 
\textbf{b}, Eye diagram SNR dependence with optical power. The eye diagrams are taken with 50\,mV $V_{\rm{pp}}$ at 3\,Gbps. The red curve is square root fitting of the experimental data.
}
\label{SFig:SNR_experiment}
\end{figure}

Here we present experimental SNR results with different drive voltage and optical power, and compare it with theoretical result Eq.\,\ref{eq:snr}. The results shown in Fig.\,\ref{SFig:SNR_experiment}a is from the eye diagrams shown in the main text Fig\,4, with 17\,dBm optical input power at 1\,Gbps data rate. When the drive voltage increases from 5\,mV to 10\,mV, 20\,mV, we see a relatively linear increase of the SNR, consistent with Eq.\,\ref{eq:snr}. The SNR also improves with increasing optical power as shown in Fig.\,\ref{SFig:SNR_experiment}b. The eye diagrams are taken with 50\,mV drive voltage at 3\,Gbps data rate. The red curve is the square root function fitting.

\section{SEOM cryogenic packaging}

\begin{figure}[htbp]
\centering
\includegraphics[trim={0cm 0cm 0cm 0cm},clip]{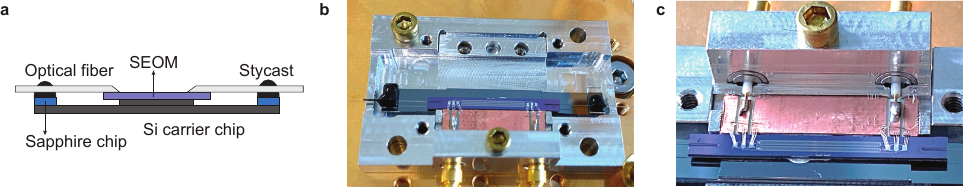}
\caption{ SEOM device packaging scheme. }
\label{SFig:package}
\end{figure}

The photonic packaging employs on-chip optical grating couplers glued with angled-facet bare fibers \cite{McKenna2019}. The grating couplers follow the design principle outlined \cite{Lomonte2021}. The SEOM chip is first glued on a Si carrier chip. The optical fibers are the aligned to the on-chip grating couplers and then glued to the SEOM chip surface using UV glue. On each side, optical fibers are anchored to two small sapphire pieces using stycast to improve the mechanical robustness, as shown in Fig.\,\ref{SFig:package}a. We achieve a 6\,dB/facet coupling efficiency after cooling to cryogenic temperature.

The fiber-connectorized SEOM is then placed in an aluminum housing to establish the interface with a PCB board. The PCB board features two  50\,$\Omega$ microwave transmission lines. The SEOM chip is positioned alongside the PCB board, and short wires are utilized for wire bonding (Fig.\,\ref{SFig:package}c). The PCB board is soldered with two GPPO microwave connectors threaded within the aluminum piece, as shown in Fig.\,\ref{SFig:package}b. There is another aluminum cover on the package to ensure that the SEOM device is well shielded with robust mechanical protection.

\section{Residual photorefractive effect and optical loss} \label{sc:opticalloss}

It has been previously reported that PECVD silicon dioxide cladding on LNOI optical waveguide causes significant photorefractive effect \cite{Xu2021a}. In our device fabrication process, we employ spin-on and thermally annealed HSQ as cladding to mitigate the photorefractive effect. We find that the photorefractive effect with this cladding approach is reduced and comparable with that of air-cladded devices. Furthermore, we identify another factor contributing to photorefractive effect and material damage in LNOI optical devices, namely the electron beam dose during lithography. This dose induces excessive optical loss, which emerges as the primary limiting factor in our SEOM devices. Finally, we present the specific impact of photorefractive effect at cryogenic temperatures for traveling-wave devices.

\begin{figure}[htbp]
\centering
\includegraphics[trim={0cm 0cm 0cm 0cm},clip]{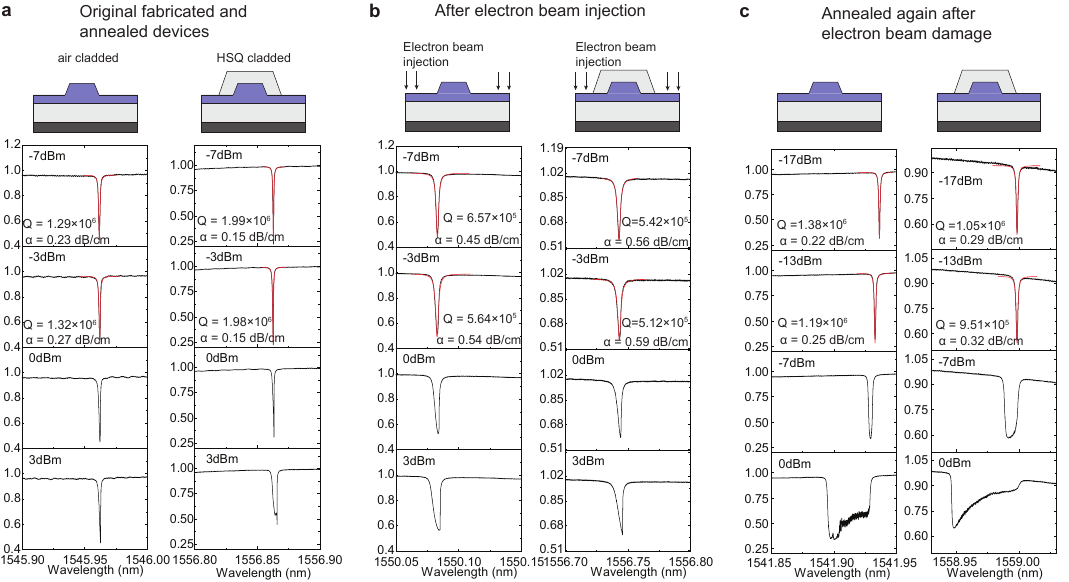}
\caption{ Evaluation of photorefractive effect and optical loss in micro-ring resonators. The wavelength is scanned from long wavelength to short wavelength at 1\,nm/s. The labeled optical input powers refer to the power in the bus waveguide leading to the micro-ring resonators. Measured optical quality factor and the corresponding propagation loss are labeled. \textbf{a}, The original device transmission after TFLN etching and thermal annealing. \textbf{b}, Device transmission after electron beam exposure. \textbf{c}, Remeasured device transmission after a 2nd thermal annealing at 300\,$^{\circ}$C for 1\,h. }
\label{SFig:PR_loss}
\end{figure}

\subsection{Photorefractive effect of HSQ cladded LNOI devices}

We fabricate micro-ring resonators to quantitatively assess the photorefractive effect and process-induced optical loss. The waveguide etching and HSQ cladding follow the procedures outlined in section \ref{sc:fab}, with selected devices omitting the HSQ cladding (thus air-cladded), serving as references. After the device fabrication, thermal annealing is required to mitigate the material absorption and restore device transmission \cite{Shams-Ansari2022}. Here we perform annealing at 400\,$^{\circ}$C for 1\,h in a nitrogen atmosphere. Immediately following the annealing, we measure the quality factor and characterize the photorefractive effect in the ring resonator as shown in Fig.\,\ref{SFig:PR_loss}a. The laser scan speed for the measurement is set at 1\,nm/s, scanning from long to short wavelength, to better reveal the photorefractive effect through resonance shift \cite{Xu2021a}. The reference air-cladded device exhibits no discernible resonance shift even at a high input power of 3\,dBm, maintaining a quality factor of 1.3\,million. For the HSQ cladded device, the optical quality factor is higher compared to the air-cladded resonators, as the cladding layer reduces the index contrast of the sidewall roughness. The measured resonance exhibits a quality factor close to 2\,million. The photorefractive effect in the HSQ-cladded device is not noticeable up to 0\,dBm optical power in the bus waveguide, indicating significant improvement over PECVD cladded devices \cite{Xu2021a}. At 3\,dBm optical power, the observed resonance distortion suggests that thermal effect dominates over photorefractive effect. Therefore, the HSQ cladding layer contributes to lower optical loss and a weak photorefractive effect comparable with that of the air-cladded devices.

\subsection{Electron beam induced material damage in LNOI}

In the process flow shown in section \ref{sc:fab}, we have observed that the SEOM devices after metal deposition experience a decrease in transmission and an increase in propagation loss (from 0.2\,dB/cm to 0.8\,dB/cm). After ruling out the metal-induced absorption loss as the cause, we pin down the actual cause of loss is from the electron beam injected into TFLN during the electron beam lithography. We confirm this by controlled ebeam exposure test on micro-ring resonators. After the initial annealing, we use the electron beam writer to expose selected areas near the optical waveguide (only electron beam is injected, no resist spinning or any other following processes). The exposed area is 2\,$\mathrm{\mu m}$ away from the optical waveguide edge and the electron beam dose is 600\,$\mathrm{\mu C/cm^2}$, similar to the dosage of the SEOM device fabrication process. Although the exposed area is not directly on the optical waveguide, the scattered electrons still manage to  expose the waveguide regime through proximity effect.

After the electron beam exposure, a noticeable decrease of the quality factor of the ring resonators is observed, as shown in Fig\,\ref{SFig:PR_loss}b. The air-cladded and HSQ-cladded ring resonators possess quality factor of 1.3\,million and 2.0\,million, respectively, prior to the electron beam exposure. However, after the exposure, the quality factor drops to around 600\,thousand and 500\,thousand, respectively. The resonance shift in resonance scanning remain relatively weak after the electron beam exposure. As a potential remedy, thermal annealing is performed again in attempt to recover the quality factor, and the result is shown in Fig.\,\ref{SFig:PR_loss}c. The device is annealed at 300\,$^{\circ}$C for 1\,h. The air-cladded device can recover its quality factor to its pre-ebeam value, while the HSQ-cladded devices exhibit partial recovery. If annealed for a longer time or at a higher temperature, the quality factor can also be fully recovered. However this second thermal annealing presents two major challenges. Firstly, it is incompatible with our SEOM fabrication flow, as the on-chip niobium is prone to oxidation during the annealing process. Even at low temperature 200\,$^{\circ}$C annealing under vacuum (1\,$\times 10^{-5}$ torr at 200\,$^{\circ}$C), the superconductive transition temperature decreases from 8\,K to around 2\,K. Secondly, the photorefractive effect induced resonance shift is significantly amplified after the 2nd annealing as indicated by the high-optical-power resonance sweep in Fig,\,\ref{SFig:PR_loss}c. The underlying mechanism behind the enhanced resonance shift after electron beam inject followed by thermal annealing is subject to future studies. The impact of this residual photorefractive effect on our SEOM device at cryogenic temperature will be discussed in the subsequent section.

Given this incompatibility challenge between superconductor and thermal annealing, our current SEOM devices experience an elevated optical loss (0.8\,dB/cm) induced by electron beam exposure. This damage is solely caused by the electron beam injection during the ebeam lithography process. Switching to photolithography using a high resolution stepper will eliminate this problem \cite{Luke2020, Liu2021}. Additionally, alternative solutions includes annealing in ultra-high vacuum chambers to protect the superconductivity, or adopting alternative fabrication processes that allow thermal annealing after electron beam lithography but before the superconductor deposition. One example of such a process is the use of a high-temperature tolerant liftoff resist. These approaches provide potential strategies to mitigate the electron beam-induced optical loss and enhance the performance of our SEOM devices.

\subsection{The impact of photorefractive effect at cryogenic temperatures}

In the previous section, we have shown that the photorefractive effect in LNOI devices can vary significantly depending on the fabrication process. Here we present how the photorefractive effect specifically impacts our SEOM device at cryogenic temperatures. The SEOM device we use is based on a Mach-Zehnder interferometer (MZI). In the presence of a strong photofrefractive effect, the light induced material index change can lead to phase drift in the two arms of the interferormeter, affecting the output power. To illustrate this, we compare two devices processed differently: One device (Device 2) undergoes waveguide etching, HSQ cladding and subsequent thermal annealing, and the other device (Device 1) follows through the same process first and then is deposited with metal through liftoff and further thermally annealed. As shown in Fig.\,\ref{SFig:PR_loss}a and c, the photorefractive effect induced index change in Device 1 is considerably stronger than that in Device 2.

\begin{figure}[htbp]
\centering
\includegraphics[trim={0cm 0cm 0cm 0cm},clip]{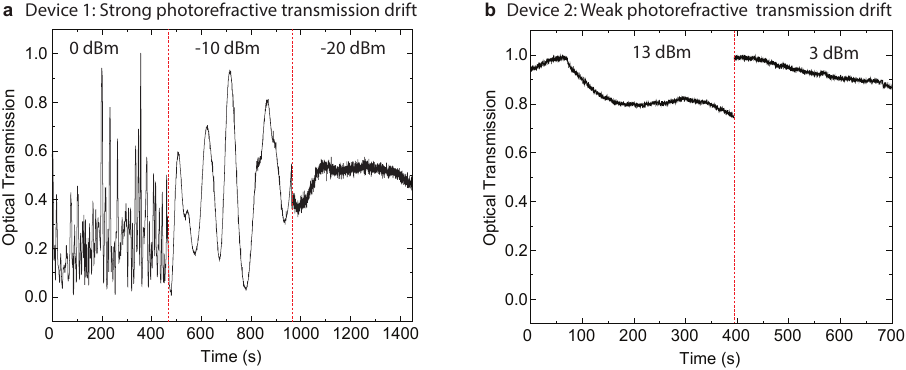}
\caption{ Optical transmission drift of two MZI-type SEOM devices at cryogenic temperature caused by the photorefractive effect. }
\label{SFig:drift}
\end{figure}

The optical transmission drift of these two devices at 4\,K are shown in Fig.\,\ref{SFig:drift}. In the case of Device 2, with laser powers of 13\,dBm and 3\,dBm, and the drift is relatively slow, within 5\% over a few minutes. However, for Device 1, strong optical transmission drift occurs even with 0\,dBm laser input. The drifting speed decreases as optical power decreases, and the transmission remains relatively stable only when the input power intensity is down to -20\,dBm. This transmission drift is of particular concern for applications that require stable operation of the SEOM device, such as digital data links where a stable transmission bias point is critical. With the slow transmission drift observed in Device 2, the stability is sufficient for experimental demonstrations. However, for long-term stability, active feedback on the SEOM or laser wavelength is necessary. Another potential solution is to utilize advanced modulation formats that do not rely on the absolute phase stability, such as phase modulator in differential phase shift keying (DPSK). Despite the relatively fast drift observed in Device 1, the drifting speed is well below kHz, whose impact can be compensated for data transmission in the GHz speed range.

\section{SEOM device performance projection} \label{sc:projection}

In sections \ref{sc:theory}, \ref{sc:EOresponse} and \ref{sc:snr}, we present how the electro-optic transduction efficiency, electro-optic bandwidth and the eye diagram SNR are related to the SEOM device parameters. In this section we discuss the current limiting factors and provide projections for future SEOM devices. We focus on two essential figure of merits for electro-optic modulators: the transduction efficiency (or modulation efficiency) and the bandwidth. According to Eq.\,\ref{eq:solution}, the optical loss of the electro-optic materials sets the fundamental limit on the transduction efficiency by determining the maximum achievable modulator length. On the other hand, the microwave loss imposes the fundamental limit on the electro-optic bandwidth under the velocity-matched condition. For the completeness of the discussion, Fig.\,5 in the main text is presented here again.

\begin{figure}[htbp]
\centering
\includegraphics[trim={0cm 0cm 0cm 0cm},clip]{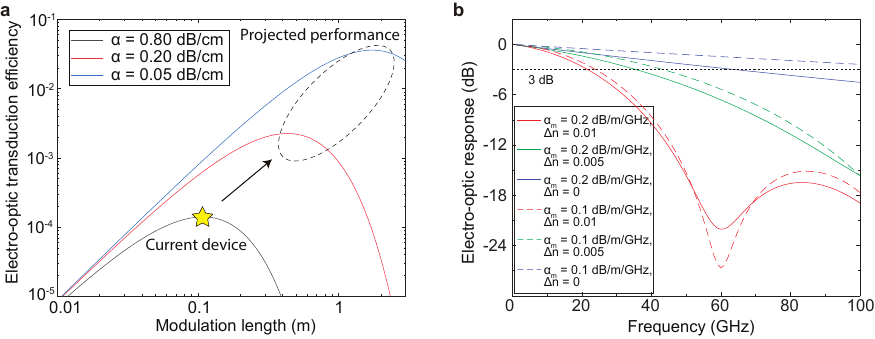}
\caption{\textbf{a}, Numerical simulation of the SEOM electro-optic transduction efficiency with varying modulation length in each arm. The simulation considers an optical input of 10\,dBm and a microwave frequency of 10\,GHz are used. \textbf{b}, Numerical simulation of electro-optic response of an SEOM with a  modulation length of 0.5\,m. }
\label{SFig:projection}
\end{figure}

As shown in Eq.\,\ref{eq:efficiency}, there exists an optimal modulation length $l = 2/\alpha_o$, that maximizes the transduction efficiency. Substituting this value back into Eq.\,\ref{eq:efficiency}, we find that $\eta \propto (1/\alpha_o)^2$. However, as discussed in section \ref{sc:opticalloss}, our current fabrication process results in elevated optical loss, which is the primary limiting factor as shown in Fig.\,\ref{SFig:projection}a. With the current propagation loss of 0.8\,dB/cm, the optimal modulation length is limited to 0.1\,m (in each interference arm of the MZI modulator). In section \ref{sc:opticalloss} we proposed solutions to reduce the propagation loss back to 0.2\,dB/cm. With this propagation loss, the optimal modulation length would increase to 0.5\,m and a transduction efficiency above 0.1\,\% could be achieved. Also, even lower propagation loss down to 0.027\,dB/cm has been reported for TFLN \cite{Zhang2017b}. With a 0.05\,dB/cm propagation loss, the transduction efficiency will approach 10\% at meter-long modulation range.

The electro-optic frequency response is captured by the last term in Eq.\,\ref{eq:solution_B} and \ref{eq:solution_R}. Assuming perfect velocity matching, the bandwidth limit arises from the microwave loss. Fig.\,\ref{SFig:projection}b illustrates the electro-optic response with different index mismatch and microwave losses. In this analysis, we assume the microwave propagation loss (in dB/m) is linearly proportional to the frequency, by a linear coefficient $\alpha_m$ (dB/m/GHz). This assumption is equivalent to considering a constant frequency-quality-factor product $fQ$ for superconducting microwave resonators. For microwave losses of $\alpha_m=$0.2 and 0.1\,dB/m/GHz, the corresponding $fQ$ products are $8\times 10^{11}$ and $1.6\times 10^{12}$, respectively. It should be noted that much higher $fQ$ product than these on TFLN has been reported ($1\times 10^{13}$ \cite{Xu2021b}). With this achievable microwave loss or $fQ$ product, the electro-optic bandwidth would be limited by the optics and microwave index mismatch. For a modulation length of 0.5\,m, if the index difference is 0.01, the bandwidth is estimated to be 20\,GHz. By employing thermal fine tuning of the velocity of the superconducting microwave, the index mismatch can be reduced further to below 0.005. As shown in Fig.\,\ref{SFig:projection}b, this can result in an electro-optic bandwidth approaching 50 to 100\,GHz.

As stated in the main text, we use a 0.1\,m-long each arm SEOM with 20\,dB total optical insertion loss for optical readout of an RSFQ circuit. In the cryogenic environment, we applied a laser power of 17\,dBm to the device and performed the optical readout at 1\,Gbps. If the limit on the optical loss is lifted (0.05\,dB/cm) and assuming a 2.5\,dB/facet coupling loss, a 0.5\,m-long SEOM in each arm can have a total insertion loss of 10\,dB and a $V_\pi$ that is 10 times lower compared with the 0.1\,m-long device. According to Eq.\,\ref{eq:snr}, this would enable optical readout of the RSFQ circuit at 10\,Gbps with an optical input power of 0\,dBm (100\,fJ/bit). If we can enhance the receiver sensitivity close to its theoretical limit (the current experimental SNR is 10\,dB lower than that of the theoretical limit as shown in Sec.\,\ref{sc:snr}), the input power requirement could be further reduced to -10\,dBm (10\,fJ/bit).

\bibliography{sup_references, References}